\documentstyle[12pt]{article}

\newcommand{\eps}{\varepsilon}
\newcommand{\arth}{{\rm arth}}

\def\N{{\rm I\kern-.1567em N}}

\def\R{{\rm I\kern-.1567em R}}
\def\C{{\rm C\kern-4.7pt
\vrule height 7.7pt width 0.4pt depth -0.5pt \phantom {.}}}
\def\Z{{\sf Z\kern-4.5pt Z}}

\def\eps{\varepsilon}

\begin{document}

\newtheorem{theorem}{Theorem}[section]
\renewcommand{\thetheorem}{\arabic{section}.\arabic{theorem}}
\newtheorem{definition}[theorem]{Definition}
\newtheorem{deflem}[theorem]{Definition and Lemma}
\newtheorem{lemma}[theorem]{Lemma}
\newtheorem{example}[theorem]{Example}
\newtheorem{remark}[theorem]{Remark}
\newtheorem{remarks}[theorem]{Remarks}
\newtheorem{cor}[theorem]{Corollary}
\newtheorem{pro}[theorem]{Proposition}
\newtheorem{proposition}[theorem]{Proposition}

\renewcommand{\theequation}{\thesection.\arabic{equation}}

\begin{titlepage}
\vspace*{3cm}
\begin{center}
{\Large\bf Slow Motion of  Charges Interacting Through the \bigskip
\\Maxwell Field}\\
\vspace{3cm}
{\large Markus Kunze}\medskip\\
        Mathematisches Institut der Universit\"at K\"oln\\
        Weyertal 86, D-50931 K\"oln, Germany\\
        email: mkunze@mi.uni-koeln.de\bigskip\\
{\large Herbert Spohn}\medskip\\
        Zentrum Mathematik and Physik Department, TU M\"unchen\\
        D-80290 M\"unchen, Germany\\
        email: spohn@mathematik.tu-muenchen.de\bigskip\\
\end{center}
\date{\today}
\vspace{3cm}

\begin{abstract}\noindent
We study the Abraham model for $N$ charges
interacting with the Maxwell field. On the scale of the charge
diameter, $R_{\varphi}$, the charges are a distance $\eps^{-1}R_{\varphi}$
apart and have a velocity $\sqrt{\eps} c$ with $\eps$ a small
dimensionless parameter. We follow the motion of the charges over
times of the order $\eps^{-3/2}R_{\varphi}/c$ and prove that on this
time  scale their motion is well approximated by the Darwin
Lagrangian. The mass is renormalized. The interaction is dominated
by the instantaneous Coulomb forces, which are of the order $\eps^{2}$.
The magnetic fields and first order retardation
generate the Darwin correction of the order $\eps^{3}$.
Radiation damping would be of the order $\eps^{7/2}$.
\end{abstract}

\end{titlepage}


\section{Introduction}

Classical charges interact through Coulomb forces, as one learns
in every course on electromagnetism. Presumably the best realization
in nature is a strongly ionized gas, for which the Darwin correction to
the Coulomb forces is of importance, since under standard conditions
the velocities cannot be considered small as compared
to the velocity of light, cf.~\cite[\S 65]{LL}.
Thus, given $N$ charges, with positions $r_\alpha$,
velocities $u_\alpha$, charges $e_\alpha$,
and masses $m_\alpha$, $\alpha=1, \ldots, N$, their motion is governed
by the Lagrangian
\begin{eqnarray}\label{LCD}
   {\cal L}_{{\rm D}} & = & \sum_{\alpha=1}^N \Big(\frac{1}{2}
   m_\alpha u_\alpha^2+\frac{1}{8c^2}m_\alpha^\ast
   u_\alpha^4\Big)-\frac{1}{2}
   \sum_{\stackrel{\alpha, \beta=1}{\alpha\neq\beta}}^N
   \frac{e_\alpha e_\beta}{4\pi |r_\alpha-r_\beta|}\nonumber
   \\ & & +\frac{1}{4c^2}\sum_{\stackrel{\alpha, \beta=1}{\alpha\neq\beta}}^N
   \frac{e_\alpha e_\beta}{4\pi |r_\alpha-r_\beta|}
   \Big(u_\alpha\cdot u_\beta + {|r_\alpha-r_\beta|}^{-2}
   (u_\alpha\cdot [r_\alpha-r_\beta])(u_\beta\cdot [r_\alpha-r_\beta])\Big),
\end{eqnarray}
$c$ denoting the velocity of light. The first term is the kinetic
energy with a $u_\alpha^4$-correction of a strength $m_\alpha^\ast$
depending on the precise model ($m_\alpha^\ast=m_\alpha$ for a relativistic
particle). The second term is the Coulomb potential, whereas
the third term is the Darwin potential, which decays as the
Coulomb potential and has a velocity dependent strength.

On a more fundamental level, the forces between the charges are
mediated through the electromagnetic field. The instantaneous
Coulomb-Darwin interaction is a derived concept only. To understand the
emergence of such an interaction, in this paper we
will investigate the coupled system, charges and Maxwell field,
and we will prove that in a certain limit the motion of the
charges is well approximated by the Lagrange equations for
${\cal L}_{{\rm D}}$.

Let us first describe how the charges are coupled to the Maxwell
field. To avoid short-distance singularities, we assume that
the charge is spread out over a distance $R_\varphi$, which
physically is of order of the classical electron radius.
Thus charge $\alpha$ has a charge distribution $\rho_\alpha$
which for simplicity we take to be of the form
\[ \rho_\alpha(x)=e_\alpha\varphi(x),\quad x\in\R^3, \]
where the form factor $\varphi$ satisfies
$$ 0\le\varphi\in C_0^\infty(\R^3)\,,\quad\varphi(x)=\varphi_r(|x|)\,,\quad
\varphi(x)=0\,\,\,\,\,\mbox{for}\,\,\,\,\,|x|\ge R_{\varphi}\,. \eqno{(C)} $$
To distinguish the true solution from the approximation
(\ref{LCD}), the position of a charge $\alpha$ in the coupled system
is denoted by $q_\alpha$ and its velocity by $v_\alpha$,
$\alpha=1, \ldots, N$. The charges then generate
the charge distribution $\rho$ and the current $j$ given by
\begin{equation}\label{rho-j-def}
   \rho(x, t)=\sum_{\alpha=1}^N\rho_\alpha(x-q_\alpha(t))
   \quad\mbox{and}\quad
   j(x, t)=\sum_{\alpha=1}^N\rho_\alpha(x-q_\alpha(t))v_\alpha(t),
\end{equation}
which satisfy charge conservation by fiat. The Maxwell field,
consisting of the electric field $E$ and the magnetic field $B$,
evolves according to
\begin{equation}\label{system0}
   c^{-1}\frac{\partial}{\partial t}  B(x, t)= -\nabla \wedge E(x, t), \quad
   c^{-1}\frac{\partial}{\partial t} E(x, t)= \nabla \wedge B(x, t)
   - c^{-1} j(x, t)
\end{equation}
with the constraints
\begin{equation}\label{system1}
   \nabla\cdot E(x, t)=\rho(x, t),\quad \nabla \cdot B(x, t)=0.
\end{equation}
The charges generate the electromagnetic field which in turn
determines the forces on the charges through the Lorentz force
equation
\begin{equation}\label{system2}
   \frac{d}{dt}\Big(m_{{\rm b}\alpha}\gamma_\alpha v_\alpha(t)\Big)
   =\int d^3x\,\rho_\alpha (x-q_\alpha(t))
   \Big[E(x,t)+v_\alpha(t)\wedge B(x, t)\Big],\quad t\in\R,
\end{equation}
for $\alpha=1, \ldots, N$. Here $m_{{\rm b}\alpha}$ is the
bare mass of charge $\alpha$ and $\gamma_\alpha$ the relativistic
factor $\gamma_\alpha={(1-v^2_\alpha/c^2)}^{-1/2}$, which
ensures $|v_\alpha|<c$. Note that there are no direct forces
acting
between the particles.
Eqns.~(\ref{rho-j-def})--(\ref{system2})
are known as Abraham model for $N$ charges.

We define the energy function by
\begin{equation}\label{energ-def}
   {\cal H}(E, B, \vec{q}, \vec{v})
   =\sum_{\alpha=1}^N m_{{\rm b}\alpha}\gamma_\alpha
   +\frac{1}{2}\,\int d^3x\,[E^2(x)+B^2(x)],
\end{equation}
with $\vec{q}=(q_1, \ldots, q_N)$ and $\vec{v}=(v_1, \ldots, v_N)$.
It then may be seen that the initial value problem corresponding to
(\ref{rho-j-def})--(\ref{system2}) has a unique weak solution
of finite energy and that ${\cal H}$ is conserved by this solution,
compare with \cite{KS} for the case of a single particle.

We assume that initially the particles are very far apart on the
scale set by $R_\varphi$. Thus we require, for $\alpha\neq\beta$,
that
\begin{equation}\label{q-ini}
  |q_\alpha(0)-q_\beta(0)|\cong \eps^{-1} R_\varphi
\end{equation}
with $\eps>0$ small. If particles would come together as close as
$R_\varphi$, our equations of motion are not trustworthy anyhow.
In addition, we require that the initial velocities be small
compared to the speed of light,
\begin{equation}\label{v-ini}
   |v_\alpha(0)|\cong\sqrt{\eps} c.
\end{equation}
Subject to these restrictions, in essence, the initial
electromagnetic field is chosen such as to minimize the energy
function ${\cal H}$ from (\ref{energ-def}), cf.~Section \ref{v-bound-sect}
for precise statements and estimates. With these initial conditions,
for the particles to travel a distance of order $\eps^{-1}R_\varphi$
it will take a time of order $\eps^{-3/2}R_\varphi/c$,
which will be the time scale of interest.
Thus physically we consider slow particles that are far apart,
and we want to follow their motion over long times.

Next note that it takes a time of order $\eps^{-1}R_\varphi/c$ for a
signal to travel between the particles. This means that on the
time scale of interest, retardation effects are small. If particles
interact through Coulomb forces, as will have to be proved, the
strength of the forces is of order $\eps^2$ since the distance
is of order $\eps^{-1}R_\varphi$. Followed over a time span
$\eps^{-3/2}R_\varphi/c$, this yields a change in velocity of order
$\sqrt{\eps}c$. On this basis we expect the orders of magnitude
(\ref{q-ini}) and (\ref{v-ini}) to remain valid over times of
order $\eps^{-3/2}R_\varphi/c$. There is one subtle point here, however.
The self-interaction of a charge with the fields renormalizes its mass.
Thus in (\ref{LCD}) the quantity $m_\alpha$ cannot be the bare mass of the
charge, the electromagnetic mass has to be added.

In theoretical physics it is common practice to count the
post-Coulombian corrections in orders of $v/c$ relative to the
motion through pure Coulomb forces. Thus the Darwin term is the
first correction and of order $(v/c)^2$. The next correction is
of order $(v/c)^3$ and accounts for damping through radiation. If
we push the Taylor expansion in Section 3 one term further, one
obtains
\begin{equation}\label{LD-dyn}
   \frac{d}{dt}\bigg(\frac{\partial {\cal L}_{{\rm D}}}
   {\partial u_\alpha}\bigg)=\frac{\partial {\cal L}_{{\rm D}}}
   {\partial r_\alpha}+ (e_{\alpha}/\,6 \pi c^{3})\sum^N_{\beta =1} 
   e_{\beta}\ddot{v}_{\beta},
\end{equation}
$\alpha=1, \ldots, N$. The physical solution has to be on the center manifold
for (\ref{LD-dyn}). At the present level of precision it suffices to
substitute the Hamiltonian dynamics to lowest order, which yields
\newpage
\begin{eqnarray*}
   &&\frac{d}{dt}\bigg(\frac{\partial {\cal L}_{{\rm D}}}
   {\partial u_\alpha}\bigg)  =  \frac{\partial {\cal L}_{{\rm D}}}
   {\partial r_\alpha}\nonumber \\&&  +\frac{e_{\alpha}}{6 \pi c^{3}}\frac{1}{2}
   \sum_{\stackrel{\beta,\beta'=1}{\beta\neq\beta'}}^N
   \big(\frac{e_{\beta}}{m_{\beta}} - 
   \frac{e_{\beta'}}{m_{\beta'}}\big)
   \frac{e_\beta e_{\beta'}}{4\pi |r_\beta-r_{\beta'}|^{3}}\Big(
   (u_\beta - u_{\beta'}) - 3 \frac{(r_\beta-r_{\beta'})\cdot
   (u_\beta - u_{\beta'})}{{|r_\beta-r_{\beta'}|}^{2}}
   (r_\beta-r_{\beta'})\Big).
\end{eqnarray*}
Note that if the ratio $e_{\alpha}/m_{\alpha}$ does not depend
on $\alpha$, then the radiation reaction vanishes and the
system does not emit dipole radiation. The next order correction
is $(v/c)^4$ and of Lagrangian form. It is discussed in \cite{LL} and 
\cite{Sch}.

In general relativity, there is a huge effort to obtain
corrections to the Newtonian orbits, which as a problem is similar
to the one discussed here. The most famous example is the
Hulse-Taylor binary pulsar, where two highly compact neutron stars
of roughly solar mass revolve around each other with a period of
7.8 h \cite{T}. In this case $(v/c) = 10^{-3}$. For gravitational systems
there is only quadrupole radiation which is of order  $(v/c)^5$.
To this order the theory agrees with
the observed radio signals within 0,3\%. In newly
designed experiments one expects highly improved precision which
will require corrections up to order $(v/c)^{11}$.


\setcounter{equation}{0}

\section{Main results}

We recall the initial conditions for the Abraham model
(\ref{rho-j-def})--(\ref{system2}), where we set $c=1$
throughout for simplicity. For the initial positions
$q_\alpha^0=q_\alpha(0)$ we require
\begin{equation}\label{q-ini2}
   C_1\eps^{-1}\le |q_\alpha^0-q_\beta^0|\le C_2\eps^{-1},
   \quad\alpha\ne\beta,
\end{equation}
for some constants $C_1, C_2>0$. For the initial velocities
$v_\alpha^0=v_\alpha(0)$ we assume
\begin{equation}\label{v-ini2}
   |v_\alpha^0|\le C_3\sqrt{\eps}
\end{equation}
with $C_3>0$. The initial fields are a sum over charge solitons,
\begin{equation}\label{ini-bed}
   E(x, 0)=E^0(x)=\sum_{\alpha=1}^N E_{v_\alpha^0}(x-q_\alpha^0)
   \quad\mbox{and}\quad
   B(x, 0)=B^0(x)=\sum_{\alpha=1}^N B_{v_\alpha^0}(x-q_\alpha^0).
\end{equation}
Here
\begin{equation}\label{EBv-def}
   E_v(x)=-\nabla\phi_v(x)+(v\cdot\nabla\phi_v(x))v
   \quad\mbox{and}\quad B_v(x)=-v\wedge\nabla\phi_v(x)
\end{equation}
and the Fourier transform of $\phi_v$ is given by
\begin{equation}\label{phiv-def}
   \hat{\phi}_v(k)=e\hat{\varphi}(k)/[k^2-{(k\cdot v)}^2],
\end{equation}
where it is understood that in $\phi_{v_\alpha^0}$
we have to set $e=e_\alpha$. For this
choice of data, the constraints (\ref{system1}) are satisfied
for $t=0$ and therefore for all $t$. In case $N=1$,
the particle would travel freely, $q_1(t)=q_1^0+v_1^0t$, $t\ge 0$,
and the co-moving electromagnetic fields  would maintain their form
(\ref{ini-bed}).

In spirit, the bounds (\ref{q-ini2}) and (\ref{v-ini2}) should propagate
in time and the form (\ref{ini-bed}) of the electromagnetic fields,
at least in approximation. On the other hand,
for two particles with opposite charge one particular
solution is the head on collision which violates the lower bound in
(\ref{q-ini2}). Considerably more delicate are solutions where some
particles reach infinity in finite time, \cite{M,X}.  Thus we simply
require that for given constants $C_\ast, C^\ast>0$ the bound
\begin{equation}\label{diff-bound}
   C_\ast\eps^{-1}\le \sup_{t\in [0,\,T\eps^{-3/2}]}
   |q_\alpha(t)-q_\beta(t)|\le C^\ast\eps^{-1},
   \quad\alpha\ne\beta,
\end{equation}
holds, which implicitly defines the first time, $T$, at which
(\ref{diff-bound}) is violated. In fact (\ref{diff-bound}) looks like an
uncheckable assumption. But, as to be shown, the optimal $T$ can be
computed on the basis of the approximation dynamics generated by the
Lagrangian (\ref{LCD}).

Under the assumption (\ref{diff-bound}) the velocity bound
propagates through the conservation of energy.
We define the electrostatic energy of the charge distributions as
\begin{equation}\label{Ecoul}
   {\cal E}_{{\rm stat}}
   =\sum_{\alpha=1}^N e_\alpha^2
   \bigg(\frac{1}{2}\int d^3k\,{|\hat{\varphi}(k)|}^2k^{-2}\bigg).
\end{equation}
and compute the energy (\ref{energ-def}) for the given
initial data. Then
\[ {\cal H}(0):={\cal H}(t=0)=
   \sum_{\alpha=1}^N m_{{\rm b}\alpha}\gamma(v_\alpha^0)
   +{\cal E}_{{\rm stat}}+{\cal O}(\eps) \]
with $\gamma(v)={(1-v^2)}^{-1/2}$.  We minimize
the electromagnetic field energy ${\cal H}_{{\rm f}}(t)
=\frac{1}{2}\int d^3x\,[E^2(x, t)+B^2(x, t)]$ at time $t$
for given $\rho$ and $j$, i.e., for given positions
$\vec{q}(t)$ and velocities $\vec{v}(t)$. Using (\ref{diff-bound}) it may
be shown that
\[ {\cal H}(t)\ge\sum_{\alpha=1}^N m_{{\rm b}\alpha}
   \gamma(v_\alpha(t))+{\cal E}_{{\rm stat}}
   +{\cal O}(\eps). \]
Since by energy conservation ${\cal H}(0)={\cal H}(t)$ and since
the dominant contributions
${\cal E}_{{\rm stat}}$ cancel exactly, we thus will continue
to have the bound $|v_\alpha(t)|\cong C\sqrt{\eps}$. (We refer
to Section \ref{v-bound-sect} in Appendix A for the complete
argument). Therefore
\begin{equation}\label{v-bound}
   \sup_{t\in [0,\,T\eps^{-3/2}]}|v_\alpha(t)|\le C_v\sqrt{\eps}
\end{equation}
with some constant $C_v>0$.

As a next step we solve the inhomogeneous Maxwell equations for the fields and
insert them into the Lorentz force equations. According to the
retarded part of the fields, retarded positions $q_\alpha(s)$,
$s\in [0, t]$, will show up. To control the Taylor expansion
of $q_\alpha(t)-q_\alpha(s)$ and thus of the retarded force,
including the Darwin term, we will need bounds not only
on positions and velocities, but also on $\dot{v}_\alpha$ and
$\ddot{v}_\alpha$. Implicitly they use that the true fields remain
close to the fields of the form (\ref{ini-bed}) evaluated at current
positions and velocities.
\begin{lemma}\label{esti} Let the initial data for the Abraham
model satisfy (\ref{q-ini2}), (\ref{v-ini2}), and (\ref{ini-bed}).
Moreover, assume
\begin{equation}\label{low-bound}
   C_\ast\eps^{-1}\le \sup_{t\in [0, T\eps^{-3/2}]}
   |q_\alpha(t)-q_\beta(t)|,
   \quad\alpha\ne\beta,
\end{equation}
for some $T>0$. Then there exist constants $C^\ast, C_v>0$ such that
(\ref{diff-bound}) and (\ref{v-bound}) hold.
In particular, $\sup_{t\in [0, T\eps^{-3/2}]}|v_\alpha(t)|\le\bar{v}<1$
for some $\bar{v}$. In addition, we find $C>0$
and $\bar{e}>0$ such that
\begin{equation}\label{wend}
   \sup_{t\in [0, T\eps^{-3/2}]}|\dot{v}_\alpha(t)|\le C\eps^2\quad
   \mbox{and}\quad
   \sup_{t\in [0, T\eps^{-3/2}]}|\ddot{v}_\alpha(t)|\le C\eps^{7/2}
\end{equation}
in case that $|e_\alpha|\le\bar{e}$, $\alpha=1, \ldots, N$.
In the estimates (\ref{diff-bound}), (\ref{v-bound}), and (\ref{wend}),
$C$ and $\bar{e}$ do depend only on $T$ and the bounds
for the initial data, but not on $\eps$.
\end{lemma}

The proof of this lemma is rather technical and will be given
in Appendix A. Using the bounds of Lemma \ref{esti}, we
expand the Lorentz force up to an error of order $\eps^{7/2}$,
cf.~Lemma \ref{force-esti}, which is the order of
radiation damping (the Coulomb force is order $\eps^2$ and
radiation damping a relative order $\eps^{3/2}$ smaller).
The terms up to order $\eps^3$ then can be collected in the form
of the Darwin Lagrangian (\ref{LCD}). We set
\[ m_\alpha=m_{{\rm b}\alpha}+\frac{4}{3}e_\alpha^2m_e\quad
   \mbox{and}\quad m_\alpha^\ast=m_{{\rm b}\alpha}+\frac{16}{15}
   e_\alpha^2 m_e \]
with the electromagnetic mass $m_e=\frac{1}{2}\int
d^3k\,{|\hat{f}(k)|}^2k^{-2}$ and the Darwin Lagrangian
\begin{eqnarray*}
   {\cal L}_{{\rm D}}(\vec{r}, \vec{u}) & = & \sum_{\alpha=1}^N
   \Big(\frac{1}{2}m_\alpha u_\alpha^2+\frac{\eps}{8}m_\alpha^\ast
   u_\alpha^4\Big)-\frac{1}{2}
   \sum_{\stackrel{\alpha, \beta=1}{\alpha\neq\beta}}^N
   \frac{e_\alpha e_\beta}{4\pi |r_\alpha-r_\beta|}
   \\ & & +\,\frac{\eps}{4}\sum_{\stackrel{\alpha, \beta=1}{\alpha\neq\beta}}^N
   \frac{e_\alpha e_\beta}{4\pi |r_\alpha-r_\beta|}
   \Big(u_\alpha\cdot u_\beta + {|r_\alpha-r_\beta|}^{-2}
   (u_\alpha\cdot [r_\alpha-r_\beta])(u_\beta\cdot [r_\alpha-r_\beta])\Big)
   \\ & &
\end{eqnarray*}
for $\vec{r}=(r_1, \ldots, r_N)$ and $\vec{u}=(u_1, \ldots, u_N)$.
The comparison dynamics is then
\begin{equation}\label{CD-dyn}
   \frac{d}{dt}\bigg(\frac{\partial {\cal L}_{{\rm D}}}
   {\partial u_\alpha}\bigg)=\frac{\partial {\cal L}_{{\rm D}}}
   {\partial r_\alpha},\quad\alpha=1, \ldots, N.
\end{equation}
It conserves the energy
\begin{equation}\label{HCD-eps}
   {\cal H}_{{\rm D}}(\vec{r}, \vec{u})
   =\sum_{\alpha=1}^N\Big(\frac{1}{2}m_\alpha u_\alpha^2
   +\eps\frac{3}{8} m_\alpha^\ast u_\alpha^4\Big)
   +\frac{1}{2}\sum_{\stackrel{\alpha, \beta=1}{\alpha\neq\beta}}^N
   \frac{e_\alpha e_\beta}{4\pi |r_\alpha-r_\beta|}\,.
\end{equation}
Because of the Coulomb singularity, in general the solutions to
(\ref{CD-dyn}) will exist only locally in time, the only exception
being when all charges have the same sign, in which case energy
conservation yields global existence. In the corresponding
gravitational problem, for a set of positive phase space measure,
mass can be transported to infinity in a finite time, \cite{X}.
We do not know whether this can happen also for
the Coulomb problem.

We set
\begin{equation}\label{ltb}
   q_\alpha^0=\eps^{-1}r_\alpha^0\quad\mbox{and}\quad
   v_\alpha^0=\sqrt{\eps}u_\alpha^0,\quad\alpha=1, \ldots, N,
\end{equation}
with $r_\alpha^0\neq r_\beta^0$ for $\alpha\neq\beta$.
Then (\ref{q-ini2}) and (\ref{v-ini2}) are satisfied.
During the initial time slip of order $\eps^{-1}$ the fields build up
the forces between particles and adjust to their motion. Thus during
that period the dynamics of the particles is not well approximated by
the Darwin Lagrangian and we correct the initial data of the
comparison dynamics to the true positions and velocities only at the
end of the initial time slip. To take into account that the
comparison dynamics will have no global solutions in time, in general,
we define $\tau\in ]0, \infty]$ to be the first
time when either $\lim_{t\to\tau^-}|r_\alpha(t)-r_\beta(t)|=0$ for some
$\alpha\neq\beta$ or $\lim_{t\to\tau^-}|r_\alpha(t)|=\infty$
for some $\alpha$ holds for the comparison dynamics (\ref{CD-dyn}).

As our main approximation result we state

\begin{theorem}\label{main-thm} Let $T>0$ be fixed.
Define $\tau\in ]0, \infty]$ as above and fix some $\delta_0\in ]0, \tau[$.
For the Abraham model let the initial data be given by (\ref{ltb})
and (\ref{ini-bed}). Furthermore we require $|e_\alpha|\le\bar{e}$, with
$\bar{e}=\bar{e}(T, {\rm data})>0$ from Lemma \ref{esti}. Let
$t_{0}=4(R_\varphi+C^\ast\eps^{-1})$. We adjust the initial data of the
comparison dynamics such that $q_\alpha(t_0)=\eps^{-1} r_\alpha(\eps^{3/2} t_0)$
and $v_\alpha(t_0)=\sqrt{\eps}u_\alpha(\eps^{3/2} t_0)$, $\alpha=1, \ldots, N$.

Then there exists a constant $C>0$ such that for all
$t\in [t_{0}, \min\{\tau-\delta_0, T\}\,\eps^{-3/2}]$ we have
\begin{equation}\label{com}
   |q_\alpha(t)-\eps^{-1}r_\alpha(\eps^{3/2}t)|\le C\sqrt{\eps},\quad
   |v_\alpha(t)-\sqrt{\eps}u_\alpha(\eps^{3/2}t)|\le C\eps^2,
   \quad\alpha=1, \ldots, N.
\end{equation}
\end{theorem}
{\bf Remarks} (i) If we are satisfied with the precision from the pure
Coulomb dynamics, then in (\ref{com}) we loose one power in $\eps$.
In this case, we can adjust the initial data of the comparison
dynamics at time $t=0$, and then (\ref{com}) holds for all
$t\in [0, \min\{\tau-\delta_0, T\}\,\eps^{-3/2}]$. \smallskip

\noindent (ii) In fact the initial data need not to be adjusted exactly at
$t=t_0$, a bound
\[ |q_\alpha(t_0)-\eps^{-1} r_\alpha(\eps^{3/2}t_0)|
   \sim\sqrt{\eps}\quad\mbox{and}\quad
   |v_\alpha(t_0)-\sqrt{\eps}u_\alpha(\eps^{3/2} t_0)|\sim\eps^2 \]
would be sufficient.


\setcounter{equation}{0}

\section{Self--action and mutual interaction}

In this section we expand the Lorentz force term
\begin{equation}\label{self-act}
   F_\alpha(t)=\int d^3x\,\rho_\alpha (x-q_\alpha(t))
   \big[E(x,t)+v_\alpha(t)\wedge B(x, t)\big].
\end{equation}
Since the fields $(E, B)$ are a solution to the inhomogeneous Maxwell's
equations, we may decompose them in the initial and the retarded fields,
\[ E(x,t)=E^{(0)}(x, t)+E^{(r)}(x,t)\quad\mbox{and}\quad
   B(x,t)=B^{(0)}(x, t)+B^{(r)}(x,t), \]
where
\begin{eqnarray*}
   \hat{E}^{(0)}(k, t) & = & \cos |k|t\,\hat{E}(k, 0)
   -i\,\frac{\sin |k|t}{|k|}\,k\wedge\hat{B}(k, 0), \\
   \hat{B}^{(0)}(k, t) & = & \cos |k|t\,\hat{B}(k, 0)
   +i\,\frac{\sin |k|t}{|k|}\,k\wedge\hat{E}(k, 0), \\
   \hat{E}^{(r)}(k, t) & = & -\int_0^t ds\,\cos |k|(t-s)\,\hat{j}(k, s)
   +i\,\int_0^t ds\,\frac{\sin |k|(t-s)}{|k|}\,\hat{\rho}(k, s)k, \\
   \hat{B}^{(r)}(k, t) & = & -i\,\int_0^t ds\,\frac{\sin |k|(t-s)}{|k|}\,
   k\wedge\hat{j}(k, s),
\end{eqnarray*}
cf.~\cite[Section 4]{MK-S-2}, with $j(x, t)$ and $\rho(x, t)$ from
(\ref{rho-j-def}). Accordingly we can rewrite $F_\alpha(t)$ in
(\ref{self-act}) as
\begin{eqnarray}
   F_\alpha(t) & = & \int d^3x\,\rho_\alpha(x-q_\alpha(t))[E^{(0)}(x, t)
   +v_\alpha (t)\wedge B^{(0)}(x, t)] \nonumber
   \\ & & +\,\int d^3x\,\rho_\alpha(x-q_\alpha(t))[E^{(r)}(x, t)+v_\alpha(t)
   \wedge B^{(r)}(x, t)]
   \nonumber \\ & = & F^{(0)}_\alpha (t)+F^{(r)}_\alpha(t).
   \label{zerle}
\end{eqnarray}

First we consider $F^{(0)}_\alpha (t)$.

\begin{lemma}\label{Falph0} For $t\in [t_0, T\eps^{-3/2}]$,
with $t_0=4(R_\varphi+C^\ast\eps^{-1})$, we have $F^{(0)}_\alpha (t)=0$.
\end{lemma}
{\bf Proof\,:} If $S(t)$ denotes the solution group generated by
the free wave equation in $D^{1, 2}(\R^3)\oplus L^2(\R^3)$, it
follows from (\ref{ini-bed}) through Fourier transform that
\begin{eqnarray*}
   \left(\begin{array}{c} E^{(0)}(x, t) \\ \dot{E}^{(0)}(x, t)
    \end{array}\right) = \bigg[S(t)\left(\begin{array}{c} E^{(0)}(\cdot, 0)
    \\ \dot{E}^{(0)}(\cdot, 0)\end{array}\right)\bigg](x)
    = -\,\sum_{\beta=1}^N e_\beta\int_{-\infty}^0
    ds\,[S(t-s)\Phi_E^\beta(\cdot-q^0_\beta-v^0_\beta s)](x),
\end{eqnarray*}
where $\Phi_E^\beta(x)=(\varphi(x)v^0_\beta, \nabla\varphi(x))$.
The analogous formula is valid for $B^{(0)}(x, t)$, with
$\Phi_E^\beta$ to be replaced with $\Phi_B^\beta(x)
=(0, v^0_\beta\wedge\nabla\varphi(x))$. For fixed $1\le\beta\le N$
and $x\in\R^3$ with $|x-q^0_\beta|\le t-R_\varphi$ assumption $(C)$ yields
${[S(t-s)\Phi_E^\beta(\cdot-q^0_\beta-v^0_\beta s)]}_1(x)=0$ for all $s\le 0$
by means of Kirchhoff's formula and Lemma \ref{esti}, ${[\ldots]}_1$
denoting the first component. As for $t\in [t_0, T\eps^{-3/2}]$
and $|x-q^0_\beta|>t-R_\varphi$ we obtain
\begin{eqnarray*}
   |x-q_\alpha(t)| & \ge & |x-q^0_\beta|-|q_\alpha(t)-q_\beta(t)|
   -|q_\beta(t)-q^0_\beta|\ge t-R_\varphi-C^\ast\eps^{-1}-C\sqrt{\eps}\,t
   \\ & \ge & t_0/2-R_\varphi-C^\ast\eps^{-1}\ge R_\varphi
\end{eqnarray*}
for $\eps$ small by Lemma \ref{esti}, the claim follows.
{\hfill$\Box$}\bigskip

Turning then to $F^{(r)}_\alpha (t)$ in (\ref{zerle}), we write
this term in Fourier transformed form and use (\ref{rho-j-def}) to
obtain
\begin{equation}\label{linny}
   F^{(r)}_\alpha (t)=e^2_\alpha F^{(r)}_{\alpha\alpha} (t)
   + \sum_{\stackrel{\beta=1}{\beta\neq\alpha}}^N
   e_\alpha e_\beta F^{(r)}_{\alpha\beta}(t),
\end{equation}
with
\begin{eqnarray}\label{fab}
   F_{\alpha\beta}^{(r)}(t) & = & \int_0^t ds\,
   \int dk\,{|\hat{\varphi}(k)|}^2
   e^{-ik\cdot[q_\alpha(t)-q_\beta(s)]}\,\Bigg\{
   -\cos|k|(t-s)\,v_\beta(s)+i\,\frac{\sin |k|(t-s)}{|k|}\,k
   \nonumber \\ & &  \hspace{15em} -i\,\frac{\sin
   |k|(t-s)}{|k|}\,v_\alpha(t)\wedge
   (k\wedge v_\beta(s))\Bigg\},
\end{eqnarray}
$\alpha, \beta=1, \ldots, N$. The term $F^{(r)}_{\alpha\alpha}(t)$
accounts for the self-force, whereas $F^{(r)}_{\alpha\beta}(t)$
for $\beta\neq\alpha$ represents the mutual interaction force
between particle $\alpha$ and particle $\beta$. These both contributions
are dealt with separately in the following two subsections.

Before going on to this, we state an auxiliary result.

\begin{lemma}\label{tayl} Let $1\le\alpha, \beta\le N$, $\alpha\neq\beta$.
For $t\in [t_0, T\eps^{-3/2}]$ we have
\begin{itemize}
\item[(a)] $\displaystyle -\int_0^t ds\,\int dk\,{|\hat{\varphi}(k)|}^2
   e^{-ik\cdot[q_\alpha(t)-q_\beta(s)]}\,
   \cos|k|(t-s)\,v_\beta(s)$ \\[1ex]
   $=\displaystyle -\int_0^\infty d\tau\,\int dk\,{|\hat{\varphi}(k)|}^2
   e^{-ik\cdot\xi_{\alpha\beta}}\,
   \cos|k|\tau\,\Big\{v_\beta-i\tau (k\cdot
   v_\beta)v_\beta-\tau\dot{v}_\beta\Big\} + {\cal O}(\eps^{7/2})$,
\item[(b)] $\displaystyle i\int_0^t ds\,\int dk\,{|\hat{\varphi}(k)|}^2
   e^{-ik\cdot[q_\alpha(t)-q_\beta(s)]}\,
   \frac{\sin|k|(t-s)}{|k|}k$ \\[1ex]
   $=\displaystyle i\int_0^\infty d\tau\,\int dk\,{|\hat{\varphi}(k)|}^2
   e^{-ik\cdot\xi_{\alpha\beta}}\,
   \frac{\sin|k|\tau}{|k|}k\,\bigg\{1-ik\cdot \Big[\tau v_\beta
   -\frac{1}{2}\tau^2\dot{v}_\beta\Big]-\frac{1}{2}\tau^2 {(k\cdot v_\beta)}^2
   \bigg\} + {\cal O}(\eps^{7/2})$,
\item[(c)] $\displaystyle (-i)\int_0^t ds\,\int dk\,{|\hat{\varphi}(k)|}^2
   e^{-ik\cdot[q_\alpha(t)-q_\beta(s)]}\,
   \frac{\sin|k|(t-s)}{|k|}\,v_\alpha(t)\wedge(k\wedge v_\beta(s))$ \\[1ex]
   $=\displaystyle (-i)\int_0^\infty d\tau\,\int dk\,{|\hat{\varphi}(k)|}^2
   e^{-ik\cdot\xi_{\alpha\beta}}\,
   \frac{\sin|k|\tau}{|k|}\,v_\alpha\wedge (k\wedge v_\beta)
   + {\cal O}(\eps^{7/2})$.
\end{itemize}
Here $v_\alpha=v_\alpha(t)$, etc., and
$\xi_{\alpha\beta}=q_\alpha(t)-q_\beta(t)$.
\end{lemma}
The proof is somewhat tedious and given in Appendix B.

\subsection{Self--action}

For $t\in [t_0, T\eps^{-3/2}]$ we have
\begin{eqnarray}\label{charg}
   F_{\alpha\alpha}^{(r)}(t) & = & \int_0^{\infty} d\tau\,
   \int dk\,{|\hat{\varphi}(k)|}^2
   e^{-i(k\cdot v_\alpha)\tau}\Big(1+\frac{i}{2}(k\cdot\dot{v}_\alpha)
   \tau^2\Big)\,\Bigg\{-\cos|k|\tau\,[v_\alpha-\dot{v}_\alpha\tau]
   +i\,\frac{\sin |k|\tau}{|k|}\,k
   \nonumber \\ & &  \hspace{13em}
   -i\,\frac{\sin |k|\tau}{|k|}\,v_\alpha\wedge
   (k\wedge [v_\alpha-\dot{v}_\alpha\tau])\Bigg\}+{\cal O}(\eps^{7/2}).
\end{eqnarray}
The rigorous proof of this relation is omitted since it is
very similar to the proof of Lemma \ref{tayl} given in
Appendix B. It once more relies on the fact that
we may Taylor expand
\[ q_\alpha(s)\cong q_\alpha-v_\alpha\tau+\frac{1}{2}\dot{v}_\alpha\tau^2
   +{\cal O}(\eps^{7/2}),\quad v_\alpha(s)\cong v_\alpha-\dot{v}_\alpha\tau
   +{\cal O}(\eps^{7/2}) \]
by Lemma \ref{esti}, with $q_\alpha=q_\alpha(t)$ etc.~and
$\tau=t-s$, whence
\[ e^{-ik\cdot[q_\alpha(t)-q_\alpha(s)]}\cong e^{-i(k\cdot v_\alpha)\tau}
   \,\Big(1+\frac{i}{2}(k\cdot\dot{v}_\alpha)\tau^2\Big)
   +{\cal O}(\eps^{7/2}). \]
Introducing
\[ I_p=\int_0^{\bar{t}}\,d\tau\frac{\sin (|k|\tau)}{|k|}
   e^{-i (k\cdot v_\alpha)\tau}\,\tau^p,\quad
   J_p=\int_0^{\bar{t}}\,d\tau\cos(|k|\tau)
   e^{-i (k\cdot v_\alpha)\tau}\,\tau^p,\quad p\in\N_0, \]
Equ.~(\ref{charg}) may be rewritten as
\begin{eqnarray}\label{alphalph}
   F_{\alpha\alpha}^{(r)}(t) & = & \lim_{\bar{t}\to\infty}\bigg(
   -\int dk\,{|\hat{\varphi}(k)|}^2\,
   \bigg\{v_\alpha J_0-\dot{v}_\alpha J_1+\frac{i}{2}(k\cdot\dot{v}_\alpha)
   v_\alpha J_2\bigg\} \nonumber\\ & & \hspace{2.8em}
   + \int dk\,{|\hat{\varphi}(k)|}^2\,
   \bigg\{i\,[(1-v_\alpha^2)k + (k\cdot v_\alpha)v_\alpha] I_0
         +i\,[(v_\alpha\cdot\dot{v}_\alpha)k-(k\cdot
         v_\alpha)\dot{v}_\alpha]I_1 \nonumber\\ & & \hspace{9.8em}
         -\frac{1}{2}(k\cdot\dot{v}_\alpha)[(1-v_\alpha^2)k
         +(k\cdot v_\alpha)v_\alpha]I_2\bigg\}\bigg)+{\cal O}(\eps^{7/2}),
\end{eqnarray}
since $\dot{v}_\alpha^2={\cal O}(\eps^4)$. Denote the term containing
the $J_p$ by ${\cal J}$ and the one containing the $I_p$ by ${\cal I}$.
To evaluate the limits $\bar{t}\to\infty$, we can rely on the results
from \cite[Section 4]{MK-S-2}. We first recall that
\[ \int dk\,{|\hat{\varphi}(k)|}^2\,J_0\to 0,\quad \int
   dk\,{|\hat{\varphi}(k)|}^2\,J_1\to
   -2m_{{\rm e}}\gamma_\alpha^2\quad\mbox{as}\quad
   \bar{t}\to\infty, \]
with $\gamma_\alpha={(1-v_\alpha^2)}^{-1/2}$ and $m_{{\rm e}}
=\frac{1}{2}\int dk\,{|\hat{\varphi}(k)|}^2 k^{-2}$. Moreover,
$\nabla_v J_1=-ik J_2$, and therefore
\begin{eqnarray}\label{calJ}
   {\cal J} & \to & (-2m_{{\rm e}}\gamma_\alpha^2)\dot{v}_\alpha
   +\frac{1}{2}\,\dot{v}_\alpha\cdot\nabla_v(-2m_{{\rm e}}\gamma_\alpha^2)
   \,v_\alpha=-2m_{{\rm e}}\gamma_\alpha^2\Big(\dot{v}_\alpha
   +\gamma_\alpha^2(v_\alpha\cdot\dot{v}_\alpha)v_\alpha\Big) \nonumber\\
   & = & -2m_{{\rm e}}\Big((1+v_\alpha^2)\dot{v}_\alpha
   +(v_\alpha\cdot\dot{v}_\alpha)v_\alpha\Big)+{\cal O}(\eps^4)
\end{eqnarray}
as $\bar{t}\to\infty$, the latter equality according to the
expansion $\gamma_\alpha^2=1+v_\alpha^2+{\cal O}(v_\alpha^4)
=1+v_\alpha^2+{\cal O}(\eps^2)$ and $\gamma_\alpha^4=1+{\cal O}(\eps)$.

What concerns ${\cal I}$, we know from
\cite{KKS, MK-S-2} that $\int dk\,{|\hat{\varphi}(k)|}^2\,kI_0\to 0$,
\[ \int dk\,{|\hat{\varphi}(k)|}^2\,I_0\to
   2m_{{\rm e}}{|v_\alpha|}^{-1}\arth |v_\alpha|,
   \quad \int dk\,{|\hat{\varphi}(k)|}^2\,(k\cdot\dot{v}_\alpha)k I_2
   \to -2m_{{\rm e}}\mu(v_\alpha)\dot{v}_\alpha \]
as $\bar{t}\to\infty$, where
\[ \mu (v) z = \Big(\frac{\gamma^2}{v^2}
   -|v|^{-3}\arth |v|\Big)\,z
   + \Big(\frac{\gamma^4}{v^4}(5v^2-3)
   +3|v|^{-5}\,\arth |v|\Big)\,(v\cdot z)v \]
for $|v|<1$ and $z\in\R^3$. Consequently, since $s^{-1}\arth(s)
=1+s^2/3+s^4/5+{\cal O}(s^6)$ for $s$ close to zero,
it thus follows after some calculation that
\begin{eqnarray}\label{calI}
   {\cal I} & \to &
   -(v_\alpha\cdot\dot{v}_\alpha)\nabla_v\Big(
   2m_{{\rm e}}{|v_\alpha|}^{-1}\arth |v_\alpha|\Big)
   +\dot{v}_\alpha\,v_\alpha\cdot\nabla_v\Big(
   2m_{{\rm e}}{|v_\alpha|}^{-1}\arth |v_\alpha|\Big)
   \nonumber \\ & &
   -\,\frac{1}{2}(1-v_\alpha^2)\Big(-2m_{{\rm e}}\mu(v_\alpha)
   \dot{v}_\alpha\Big)
   -\frac{1}{2}v_\alpha\,v_\alpha\cdot\Big(
   -2m_{{\rm e}}\mu(v_\alpha)\dot{v}_\alpha \Big)
   \nonumber \\ & = &
   \bigg(\frac{2}{3}+\frac{22}{15} v_\alpha^2\bigg)m_{{\rm e}}\dot{v}_\alpha
   +\frac{14}{15}m_{{\rm e}}(v_\alpha\cdot\dot{v}_\alpha)v_\alpha+{\cal O}(\eps^4).
\end{eqnarray}
Summarizing (\ref{alphalph}), (\ref{calJ}), and (\ref{calI}),
we arrive at
\begin{lemma}\label{mota} For $t\in [t_0, T\eps^{-3/2}]$
we have
\[ F_{\alpha\alpha}^{(r)}(t) =
   -\bigg(\frac{4}{3}+\frac{8}{15} v_\alpha^2\bigg)m_{{\rm e}}\dot{v}_\alpha
   -\frac{16}{15}m_{{\rm e}}(v_\alpha\cdot\dot{v}_\alpha)v_\alpha
   +{\cal O}(\eps^{7/2}). \]
\end{lemma}

\subsection{Mutual interaction}

In this section we expand $F_{\alpha\beta}^{(r)}(t)$ from
(\ref{fab}) with $\beta\neq\alpha$.
For $p\in\N_0$ we have that
\[ A_p:=\int_0^\infty d\tau\,\int dk\,{|\hat{\varphi}(k)|}^2
   e^{-ik\cdot \xi_{\alpha\beta}}\,\frac{\sin|k|\tau}{|k|}\,\tau^p
   ={(4\pi)}^{-1}\,\int\int dx dy\,\varphi(x)\varphi(y)
   {|\xi_{\alpha\beta}+x-y|}^{p-1} \]
and
\begin{eqnarray*}
   B_p & := & \int_0^\infty d\tau\,\int dk\,{|\hat{\varphi}(k)|}^2
   e^{-ik\cdot \xi_{\alpha\beta}}\,\cos(|k|\tau)\,\tau^p
   \\ & = & (-p){(4\pi)}^{-1}\,\int\int dx dy\,\varphi(x)\varphi(y)
   {|\xi_{\alpha\beta}+x-y|}^{p-2}
   =(-p)A_{p-1},
\end{eqnarray*}
as may be seen through Fourier transform. We hence obtain from
Lemma \ref{tayl} that for $\beta\neq\alpha$ and
$t\in [t_0, T\eps^{-3/2}]$
\begin{eqnarray}\label{kemp}
   F_{\alpha\beta}^{(r)}(t) & = & -v_\beta
   (v_\beta\cdot\nabla_\xi)B_1+\dot{v}_\beta B_1-\nabla_\xi A_0
   +\frac{1}{2}(\dot{v}_\beta\cdot\nabla_\xi)\nabla_\xi A_2
   -\frac{1}{2} {(v_\beta\cdot\nabla_\xi)}^2\nabla_\xi A_2 \nonumber
   \\ & & + (v_\alpha\cdot v_\beta)\nabla_\xi A_0-v_\beta
   (v_\alpha\cdot\nabla_\xi)A_0
   + {\cal O}(\eps^{7/2}),
\end{eqnarray}
taking also into account that $A_1={(4\pi)}^{-1}$, thus $\nabla_\xi
A_1=0$. As a consequence of $|\xi_{\alpha\beta}|={\cal O}(\eps^{-1})$,
cf.~Lemma \ref{esti}, of assumption $(C)$, and of Lemma \ref{esti}, it
follows that in (\ref{kemp}) we have $-\nabla_\xi A_0={\cal O}(\eps^2)$,
while all other terms are ${\cal O}(\eps^3)$. Since e.g.
\[ \bigg|(v_\alpha\cdot v_\beta)\nabla_\xi
   A_0-(v_\alpha\cdot v_\beta)
   \bigg(-\frac{\xi_{\alpha\beta}}{4\pi{|\xi_{\alpha\beta}|}^3}\bigg)\bigg|
   \le C\eps^4, \]
with an obvious similar estimate for the other terms besides
$-\nabla_\xi A_0$, we find from (\ref{kemp})
and after some calculation that for $\beta\neq\alpha$
and $t\in [t_0, T\eps^{-3/2}]$
\begin{eqnarray*}
   F_{\alpha\beta}^{(r)}(t) & = & v_\beta
   (v_\beta\cdot\nabla_\xi)\bigg(\frac{1}{4\pi |\xi_{\alpha\beta}|}\bigg)
   -\dot{v}_\beta\bigg(\frac{1}{4\pi |\xi_{\alpha\beta}|}\bigg)
   -\nabla_\xi A_0
   +\frac{1}{2}(\dot{v}_\beta\cdot\nabla_\xi)\nabla_\xi
   \bigg(\frac{|\xi_{\alpha\beta}|}{4\pi}\bigg)
   \\ & & -\frac{1}{2} {(v_\beta\cdot\nabla_\xi)}^2\nabla_\xi
   \bigg(\frac{|\xi_{\alpha\beta}|}{4\pi}\bigg)
   + (v_\alpha\cdot v_\beta)\nabla_\xi
   \bigg(\frac{1}{4\pi |\xi_{\alpha\beta}|}\bigg)
   -v_\beta (v_\alpha\cdot\nabla_\xi)
   \bigg(\frac{1}{4\pi |\xi_{\alpha\beta}|}\bigg)
   + {\cal O}(\eps^{7/2}) \\ & = & -\nabla_\xi A_0
   -\frac{1}{8\pi |\xi_{\alpha\beta}|}\,\dot{v}_\beta
   -\frac{(\dot{v}_\beta\cdot \xi_{\alpha\beta})}
   {8\pi |\xi_{\alpha\beta}|^3}\,\xi_{\alpha\beta}
   +\frac{v_\beta^2}{8\pi |\xi_{\alpha\beta}|^3}\,\xi_{\alpha\beta}
   -\frac{3{(v_\beta\cdot\xi_{\alpha\beta})}^2}{8\pi |\xi_{\alpha\beta}|^5}
   \,\xi_{\alpha\beta} \\ & & -\frac{(v_\alpha\cdot v_\beta)}
   {4\pi |\xi_{\alpha\beta}|^3}\,\xi_{\alpha\beta}
   +\frac{(v_\alpha\cdot \xi_{\alpha\beta})}{4\pi |\xi_{\alpha\beta}|^3}
   \,v_\beta + {\cal O}(\eps^{7/2}).
\end{eqnarray*}
Finally, to deal with the lowest-order term we observe that
with $\vec{n}=\xi_{\alpha\beta}/|\xi_{\alpha\beta}|$
\begin{equation}\label{mullaig}
  \bigg|\nabla_\xi A_0+\frac{\xi_{\alpha\beta}}
  {4\pi|\xi_{\alpha\beta}|^3}\bigg|
  =\frac{1}{4|\xi_{\alpha\beta}|^2}\,
  \bigg|\int\int dx dy\,\varphi(x)\varphi(y)\Bigg(\frac{\vec{n}
  +\frac{x-y}{|\xi_{\alpha\beta}|}}{\big|\vec{n}
  +\frac{x-y}{|\xi_{\alpha\beta}|}\big|^3}-\vec{n}\Bigg)\bigg|.
\end{equation}
Defining $R=(x-y)/|\xi_{\alpha\beta}|={\cal O}(\eps)$ for $|x|, |y|\le
R_\varphi$, we can expand $\psi(R)=(\vec{n}+R)/|\vec{n}+R|$
to obtain that $\psi(R)
=\vec{n}+R-3(\vec{n}\cdot R)\vec{n}+{\cal O}(\eps^2)$.
As $\int\int dx dy\,\varphi(x)
\varphi(y)(x-y)=0$, we hence conclude that the right-hand side
of (\ref{mullaig}) is ${\cal O}(\eps^4)$. Thus we can summarize
our estimates on the mutual interaction force as follows.

\begin{lemma}\label{wolf} For $\beta\neq\alpha$ and $t\in
[t_0, T\eps^{-3/2}]$ we have
\begin{eqnarray*}
   F_{\alpha\beta}^{(r)}(t) & = &
   \frac{\xi_{\alpha\beta}}{4\pi|\xi_{\alpha\beta}|^3}
   -\frac{1}{8\pi |\xi_{\alpha\beta}|}\,\dot{v}_\beta
   -\frac{(\dot{v}_\beta\cdot\xi_{\alpha\beta})}
   {8\pi |\xi_{\alpha\beta}|^3}\,\xi_{\alpha\beta}
   +\frac{v_\beta^2}{8\pi |\xi_{\alpha\beta}|^3}\,\xi_{\alpha\beta}
   -\frac{3{(v_\beta\cdot\xi_{\alpha\beta})}^2}{8\pi |\xi_{\alpha\beta}|^5}
   \,\xi_{\alpha\beta} \\ & &
   -\frac{(v_\alpha\cdot v_\beta)}{4\pi |\xi_{\alpha\beta}|^3}\,
   \xi_{\alpha\beta}+\frac{(v_\alpha\cdot \xi_{\alpha\beta})}
   {4\pi |\xi_{\alpha\beta}|^3}\,v_\beta + {\cal O}(\eps^{7/2}).
\end{eqnarray*}
\end{lemma}

\subsection{Summary of the estimates}

By (\ref{self-act}), (\ref{zerle}), and Lemma \ref{Falph0}
we find $F_\alpha(t)=F_\alpha^{(r)}(t)$ for $t\in [t_0, T\eps^{-3/2}]$.
According to (\ref{linny}) and Lemmas \ref{mota} and \ref{wolf} we hence
have obtained the following expansion of the Lorentz force in (\ref{self-act}).
For $t\in [t_0, T\eps^{-3/2}]$ we have
\begin{eqnarray}\label{ga-def}
   F_\alpha(t) & = &
   -\bigg(\frac{4}{3}+\frac{8}{15} v_\alpha^2\bigg)m_{{\rm e}}\dot{v}_\alpha
   -\frac{16}{15}m_{{\rm e}}(v_\alpha\cdot\dot{v}_\alpha)v_\alpha
   +G_\alpha(\vec{q}, \vec{v}, \vec{\dot{v}})+{\cal O}(\eps^{7/2}),
   \nonumber \\ G_\alpha(\vec{q}, \vec{v}, \vec{\dot{v}}) & = &
   \sum_{\stackrel{\beta=1}{\beta\neq\alpha}}^N \frac{e_\alpha e_\beta}{4\pi}
   \Bigg(\frac{\xi_{\alpha\beta}}{|\xi_{\alpha\beta}|^3}
   -\frac{1}{2|\xi_{\alpha\beta}|}\,\dot{v}_\beta
   -\frac{(\dot{v}_\beta\cdot \xi_{\alpha\beta})}
   {2|\xi_{\alpha\beta}|^3}\,\xi_{\alpha\beta}
   +\frac{v_\beta^2}{2|\xi_{\alpha\beta}|^3}\,\xi_{\alpha\beta}
   -\frac{3{(v_\beta\cdot\xi_{\alpha\beta})}^2}{2|\xi_{\alpha\beta}|^5}
   \,\xi_{\alpha\beta} \nonumber\\ & & \hspace{5em}
   -\frac{(v_\alpha\cdot v_\beta)}{|\xi_{\alpha\beta}|^3}\,
   \xi_{\alpha\beta}+\frac{(v_\alpha\cdot\xi_{\alpha\beta})}
   {|\xi_{\alpha\beta}|^3}\,v_\beta\Bigg),
\end{eqnarray}
where $t_0=4(R_\varphi+C^\ast\eps^{-1})$, $\xi_{\alpha\beta}=q_\alpha(t)-q_\beta(t)$,
$v_\alpha=v_\alpha(t)$, and $v_\beta=v_\beta(t)$. Due to the Lorentz equation
$\frac{d}{dt}(m_{{\rm b}\alpha}\gamma_\alpha v_\alpha)=F_\alpha(t)$,
cf.~(\ref{system2}), we finally obtain the following lemma
by calculating the right-hand side and expanding $\gamma_\alpha$.

\begin{lemma}\label{force-esti} For $t\in [t_0, T\eps^{-3/2}]$ we have
\[ M_\alpha(v_\alpha)\dot{v}_\alpha=G_\alpha(\vec{q}, \vec{v},
   \vec{\dot{v}})+{\cal O}(\eps^{7/2}),\quad 1\le\alpha\le N, \]
with $G_\alpha$ from (\ref{ga-def}) and $M_\alpha(v)$ the $(3\times 3)$-matrix
$M_\alpha(v)(z)=(m_\alpha+\frac{1}{2}m_\alpha^{\ast}v^2)z+m_\alpha^{\ast}(v\cdot z)v$
for $v, z\in\R^3$.
\end{lemma}


\setcounter{equation}{0}

\section{Proof of Theorem \ref{main-thm}}

We need to compare a solution $(q_\alpha(t), v_\alpha(t))$ of
(\ref{rho-j-def})--(\ref{system2}) with data (\ref{ltb})
to $(\tilde{r}_\alpha(t), \tilde{u}_\alpha(t))$, where we let
\begin{equation}\label{trafo}
   \tilde{r}_\alpha(t)=\eps^{-1}r_\alpha(\eps^{3/2}t),\quad
   \tilde{u}_\alpha(t)=\sqrt{\eps} u_\alpha(\eps^{3/2}t),
\end{equation}
and where the $(r_\alpha(t), u_\alpha(t))$ are the solution
to the system induced by (\ref{CD-dyn}) with data $(r_\alpha^0,
u_\alpha^0)$.

A somewhat lengthy but elementary calculation shows that
$(\tilde{r}_\alpha(t), \tilde{u}_\alpha(t))$ satisfy
\begin{equation}\label{voll-gl}
   M_\alpha(\tilde{u}_\alpha)\dot{\tilde{u}}_\alpha
   =G_\alpha(\vec{\tilde{r}}, \vec{\tilde{u}},
   \vec{\dot{\tilde{u}}}),\quad 1\le\alpha\le N,
\end{equation}
cf.~Lemma \ref{force-esti} for the notation.
Recalling that $\tau\in ]0, \infty]$ was defined to be the first
time when either $\lim_{t\to\tau^-}|r_\alpha(t)-r_\beta(t)|=0$ for some
$\alpha\neq\beta$ or $\lim_{t\to\tau^-}|r_\alpha(t)|=\infty$
for some $\alpha$ holds, we find that (\ref{voll-gl}) is valid
for $t\in [0, (\tau-\delta_0)\eps^{-3/2}]$, for any $\delta_0\in ]0, \tau[$
which we consider to be fixed throughout.
This leads to some useful estimates on the effective dynamics.

\begin{lemma} For suitable constants $C_0, C^0, C>0$ (depending on
$\tau$, $\delta_0$, and the data) we have
\begin{equation}\label{diff-bound-vo}
   C_0\eps^{-1}\le \sup_{t\in [0,\,(\tau-\delta_0)\eps^{-3/2}]}
   |\tilde{r}_\alpha(t)-\tilde{r}_\beta(t)|\le C^0\eps^{-1},
   \quad\alpha\ne\beta,
\end{equation}
and
\begin{equation}\label{v-bound-vo}
   \sup_{t\in [0,\,(\tau-\delta_0)\eps^{-3/2}]}|
   \tilde{u}_\alpha(t)|\le C\sqrt{\eps}.
\end{equation}
\end{lemma}
{\bf Proof\,:} The bounds in (\ref{diff-bound-vo})
follow from (\ref{trafo}) and the fact that
$|r_\alpha(t)-r_\beta(t)|\ge\delta_1$ and $|r_\alpha(t)|\le C$
on $[0, \tau-\delta_0]$ for some $\delta_1>0$, $C>0$, by definition of $\tau$.
Concerning (\ref{v-bound-vo}), by conservation of the energy ${\cal H}_{{\rm D}}$
from (\ref{HCD-eps}) we obtain $C\ge {\cal H}_{{\rm D}}(\vec{r}(0), \vec{u}(0))
={\cal H}_{{\rm D}}(\vec{r}(t), \vec{u}(t))\ge\frac{1}{2}m_\alpha u^2_\alpha(t)$
as long as the solution exists, in particular for $t\in [0, \tau-\delta_0]$.
{\hfill$\Box$}\bigskip

To simplify the presentation, we henceforth omit
the tilde and write $(r, u)$ instead of $(\tilde{r}, \tilde{u})$
to denote the rescaled solution. Utilizing the bounds from Lemma
\ref{esti} and from (\ref{diff-bound-vo}), (\ref{v-bound-vo}), it
may be seen after some calculation that
\begin{equation}\label{scller}
   \Big|G_\alpha(\vec{q}, \vec{v}, \vec{\dot{v}})(t)-G_\alpha(\vec{r}, \vec{u},
   \vec{\dot{u}})(t)\Big|\le
   C\sum_{\beta=1}^N\Big(\eps^3|q_\beta(t)-r_\beta(t)|
   +\eps^{5/2}|v_\beta(t)-u_\beta(t)|+\eps|\dot{v}_\beta(t)-\dot{u}_\beta(t)|\Big)
\end{equation}
for $1\le\alpha\le N$ and $t\in [0, T\eps^{-3/2}]\cap
[0, (\tau-\delta_0)\eps^{-3/2}]=[0, \min\{\tau-\delta_0, T\}\eps^{-3/2}]$.
Note that the term $\eps^3|q_\beta-r_\beta|$ appears through
comparison of $\xi_{\alpha\beta}/|\xi_{\alpha\beta}|^3$ to
$r_{\alpha\beta}/|r_{\alpha\beta}|^3$, cf.~the form of $G_\alpha$
in (\ref{ga-def}).

Next, a general $(3\times 3)$-matrix $M(v)=a(v){\rm id}+b(v\otimes v)$
has the inverse
\[ M(v)^{-1}=a(v)^{-1}{\rm id}+\frac{b}{a(v)[a(v)+bv^2]}(v\otimes v). \]
This remark shows $|{M_\alpha(v_\alpha)}^{-1}|={\cal O}(1)$ and
$|{M_\alpha(v_\alpha)}^{-1}-{M_\alpha(u_\alpha)}^{-1}|\le C\sqrt{\eps}|v_\alpha-u_\alpha|$
for $t\in [0, \min\{\tau-\delta_0, T\}\eps^{-3/2}]$. Since $|G_\alpha(\vec{q}, \vec{v},
\vec{\dot{v}})|={\cal O}(\eps^2)$ it follows from Lemma \ref{force-esti},
(\ref{voll-gl}), and (\ref{scller}) that
\[ |\dot{v}_\alpha(t)-\dot{u}_\alpha(t)|\le C\sum_{\beta=1}^N\Big(\eps^3|q_\beta(t)-r_\beta(t)|
   +\eps^{5/2}|v_\beta(t)-u_\beta(t)|+\eps|\dot{v}_\beta(t)-\dot{u}_\beta(t)|\Big)
   +{\cal O}(\eps^{7/2}) \]
for $1\le\alpha\le N$ and $t\in [t_0, \min\{\tau-\delta_0, T\}\eps^{-3/2}]$.
Summation over $\alpha$ and choosing $\eps>0$ sufficiently small this results in
\begin{equation}\label{bri}
   \sum_{\alpha=1}^N|\dot{v}_\alpha(t)-\dot{u}_\alpha(t)|
   \le C\sum_{\alpha=1}^N\Big(\eps^3|q_\alpha(t)-r_\alpha(t)|
   +\eps^{5/2}|v_\alpha(t)-u_\alpha(t)|\Big)
   +{\cal O}(\eps^{7/2})
\end{equation}
for $t\in [t_0, \min\{\tau-\delta_0, T\}\eps^{-3/2}]$.
To use this basic estimate, we write $d_\alpha(t)=q_\alpha(t)-r_\alpha(t)$
as
\[ d_\alpha(t)=d_\alpha(t_0)+(t-t_0)\dot{d}_\alpha(t_0)+\int_{t_0}^t
   (t-s)\ddot{d}_\alpha(s)\,ds,\quad
   \dot{d}_\alpha(t)=\dot{d}_\alpha(t_0)+\int_{t_0}^t\ddot{d}_\alpha(s)\,ds. \]
We then obtain for $t\in [t_0, \min\{\tau-\delta_0, T\}\eps^{-3/2}]$
from (\ref{bri}) that
\begin{eqnarray}
   D(t) & \le & D(t_0)+(t-t_0)\bar{D}(t_0)+C\eps^3\int_{t_0}^t(t-s)D(s)\,ds
   \nonumber\\ & & +C\eps^{5/2}\int_{t_0}^t(t-s)\bar{D}(s)\,ds+C\sqrt{\eps}, \label{so1} \\
   \bar{D}(t) & \le & \bar{D}(t_0)+C\eps^3\int_{t_0}^tD(s)\,ds
   +C\eps^{5/2}\int_{t_0}^t\bar{D}(s)\,ds+C\eps^2, \label{so2}
\end{eqnarray}
where
\[ D(t)=\max_{1\le\alpha\le N}\max_{s\in [t_0, t]}|d_\alpha(s)|\quad\mbox{and}
   \quad\bar{D}(t)=\max_{1\le\alpha\le N}\max_{s\in [t_0, t]}|\dot{d}_\alpha(s)|. \]
Application of Gronwall's lemma to (\ref{so2}) yields
\begin{equation}\label{DD}
    \bar{D}(t)\le C\bigg(\bar{D}(t_0)+\eps^2+\eps^3\int_{t_0}^t
   D(s)\,ds\bigg),
   \end{equation}
and utilizing this in (\ref{so1}) implies
\[ D(t)\le D(t_0)+(t-t_0)\bar{D}(t_0)+C\sqrt{\eps}+C\eps^{-1/2}(\bar{D}(t_0)
   +\eps^2)+C\eps^3\int_{t_0}^t (t-s)D(s)\,ds. \]
Finally, $(t-s)\le C\eps^{-3/2}$ yields upon a further
application of Gronwall's lemma that
\begin{equation}\label{D}
    D(t)\le C\Big(D(t_0)+\eps^{-3/2}\bar{D}(t_0)+\sqrt{\eps}\,\Big),\quad
   t\in [t_0, \min\{\tau-\delta_0, T\}\eps^{-3/2}].
   \end{equation}
By assumption $D(t_{0})=0=\bar{D}(t_{0})$. Therefore (\ref{D}) and
(\ref{DD}) imply (\ref{com}). This completes the proof
of Theorem \ref{main-thm}. {\hfill$\Box$}\bigskip


\setcounter{equation}{0}

\section{Appendix A: Proof of Lemma \ref{esti}}

This appendix concerns the proof of Lemma \ref{esti}.
We split the proof into three subsections.

\subsection{Bounding the particle distances and the velocities}\label{v-bound-sect}

We intend to use energy conservation to show (\ref{v-bound}),
and for that reason we calculate with (\ref{ini-bed}) the field energy
\begin{eqnarray*}
   {\cal H}_{{\rm F}}(0) & = & \frac{1}{2}\int d^3x\,[E^2(x, 0)+B^2(x, 0)]
   \\ & = & \frac{1}{2}\sum_{\alpha=1}^N
   \int d^3x\,[E_{v_\alpha^0}^2(x-q_\alpha^0)
   +B_{v_\alpha^0}^2(x-q_\alpha^0)]
   + \frac{1}{2}\sum_{\stackrel{\alpha, \beta=1}{\alpha\neq\beta}}^N
   \int d^3x\,\Big[E_{v_\alpha^0}(x-q_\alpha^0)\cdot
   E_{v_\beta^0}(x-q_\beta^0)
   \\ & & \hspace{25em} +B_{v_\alpha^0}(x-q_\alpha^0)\cdot
   B_{v_\beta^0}(x-q_\beta^0)\Big].
\end{eqnarray*}
According to (\ref{EBv-def}) and \cite[Section 2]{MK-S-2}
the first term equals
\[ {\cal H}_{{\rm F}}^{(1)}(0)
   =\sum_{\alpha=1}^N e_\alpha^2 \bigg(\frac{1}{2}\int
   d^3k\,{|\hat{\varphi}(k)|}^2k^{-2}\bigg)
   \bigg[\frac{1}{|v_\alpha^0|}\,\log\frac{1+|v_\alpha^0|}
   {1-|v_\alpha^0|}-1\bigg]. \]
Denoting the term in $[\ldots]$ as $\psi(|v_\alpha^0|)$, $\psi(r)$
is odd, and hence Taylor expansion implies $\psi(r)=1+{\cal O}(r^2)$
for $r$ small. Therefore (\ref{v-ini2}) yields
\[ {\cal H}_{{\rm F}}^{(1)}(0)={\cal E}_{{\rm Coul}}+{\cal O}(\eps), \]
with ${\cal E}_{{\rm Coul}}$ from (\ref{Ecoul}). To deal with the
contributions for $\alpha\neq\beta$ in the second term, we obtain
by passing to Fourier transformed form and observing (\ref{v-ini2}) that e.g.
\[ \int d^3x\,E_{v_\alpha^0}(x-q_\alpha^0)\cdot E_{v_\beta^0}(x-q_\beta^0)
   =e_\alpha e_\beta\int d^3k\,{|\hat{\varphi}(k)|}^2k^{-2}e^{ik\cdot
   (q_\alpha^0-q_\beta^0)}+{\cal O}(\eps)={\cal O}(\eps), \]
the latter with (\ref{q-ini2}) and by passing to polar
coordinates. Thus we have shown
\begin{equation}\label{sudt1}
   {\cal H}_{{\rm F}}(0)={\cal E}_{{\rm Coul}}+{\cal O}(\eps).
\end{equation}

Next we will investigate the field energy at time $t>0$. We claim
that
\begin{equation}\label{wart}
   {\cal H}_{{\rm F}}(t)=\frac{1}{2}\int d^3x\,[E^2(x, t)+B^2(x, t)]
   \ge\frac{1}{2}\int d^3x\,E^2(x, t)\ge
   -\frac{1}{2}{\Big(\rho(\cdot, t),
   \Delta^{-1}\rho(\cdot, t)\Big)}_{L^2(\R^3)}.
\end{equation}
The easiest way to see this is to introduce potentials $A$ and $\phi$,
\[ B(x, t)=\nabla\wedge A(x, t),\quad E(x, t)=-\nabla\phi(x,t)
   -F(x, t),\quad\mbox{with}\quad F(x, t)
   =\frac{\partial A}{\partial t}(x, t), \]
for the electromagnetic field. Then $\rho=\nabla\cdot E=-\Delta\phi
-\nabla\cdot F$, and the estimate in (\ref{wart}) follows by
passing to Fourier transformed form. On the other hand,
substituting $\rho$ from (\ref{rho-j-def}) into
\[ -\frac{1}{2}{\Big(\rho(\cdot, t),
   \Delta^{-1}\rho(\cdot, t)\Big)}_{L^2(\R^3)}
   =\frac{1}{2}\int d^3k\,{|\hat{\rho}(k, t)|}^2k^{-2}, \]
by assumption (\ref{low-bound}) we can argue exactly as before
to show that the terms with $\alpha\neq\beta$ are ${\cal O}(\eps)$,
and thus
\begin{equation}\label{sudt2}
   {\cal H}_{{\rm F}}(t)\ge \frac{1}{2}\sum_{\alpha=1}^N
   \int d^3k\,{|\hat{\rho}_\alpha(k)|}^2k^{-2}+{\cal O}(\eps)
   ={\cal E}_{{\rm Coul}}+{\cal O}(\eps)
\end{equation}
for $t\in [0, T\eps^{-3/2}]$.
Consequently for $t\in [0, T\eps^{-3/2}]$ by energy conservation,
cf.~(\ref{energ-def}), by (\ref{sudt1}) and (\ref{sudt2})
\begin{eqnarray*}
   \sum_{\alpha=1}^N m_{{\rm b}\alpha}\gamma(v_\alpha^0)
   + {\cal E}_{{\rm Coul}} + {\cal O}(\eps) & = &
   \sum_{\alpha=1}^N m_{{\rm b}\alpha}\gamma(v_\alpha^0)
   + {\cal H}_F(0)
   =\sum_{\alpha=1}^N m_{{\rm b}\alpha}\gamma(v_\alpha(t))
   + {\cal H}_F(t)
   \\ & \ge & \sum_{\alpha=1}^N m_{{\rm b}\alpha}\gamma(v_\alpha(t))
   + {\cal E}_{{\rm Coul}} + {\cal O}(\eps),
\end{eqnarray*}
with $\gamma(v)={(1-v^2)}^{-1/2}$. Thus
\begin{equation}\label{brutu}
   \sum_{\alpha=1}^N m_{{\rm b}\alpha}\gamma(v_\alpha^0)
   +C\eps\ge\sum_{\alpha=1}^N m_{{\rm b}\alpha}
   \gamma(v_\alpha(t)),\quad t\in [0, T\eps^{-3/2}]
\end{equation}
with some constant $C$ depending on $C_1, C_3, C_\ast, T$.
This estimate now allows to prove (\ref{v-bound}).
Define
\[ I_+=\{\alpha\in\{1, \ldots, N\}: \gamma(v_\alpha(t))\le
   \gamma(v_\alpha^0)\}\quad\mbox{and}\quad
   I_-=\{\alpha\in\{1, \ldots, N\}: \gamma(v_\alpha(t))
   >\gamma(v_\alpha^0)\}. \]
For $\alpha\in I_+$ we have $|v_\alpha(t)|\le |v_\alpha^0|
\le C_3\sqrt{\eps}$ by (\ref{v-ini2}). Thus for $\eps$ so small
that $C_3^2\eps\le 1/2$, $\gamma(v_\alpha^0)-\gamma (v_\alpha(t))
\le\sqrt{2}|(v_\alpha^0)^2-(v_\alpha(t))^2|\le C\eps$. Therefore
by (\ref{brutu})
\[ C\eps\ge\sum_{\alpha\in I_-} m_{{\rm b}\alpha}
   \Big(\gamma(v_\alpha(t))-\gamma(v_\alpha^0)\Big). \]
Since $m_{{\rm b}\alpha}>0$ we deduce that
\[ \gamma(v_\alpha(t))\le\gamma(v_\alpha^0)+C\eps,\quad\alpha\in I_-, \]
and according to $|v_\alpha^0|\le C_3\sqrt{\eps}$ it then follows
that $|v_\alpha(t)|\le C\sqrt{\eps}$ also for $\alpha\in I_-$.
This concludes the proof of (\ref{v-bound}).

Using (\ref{q-ini2}) and (\ref{v-bound}) it is finally easy to
derive the upper bound in (\ref{diff-bound}), since
for $t\in [0, T\eps^{-3/2}]$ we have
\[ |q_\alpha(t)-q_\beta(t)|\le |q_\alpha^0-q_\beta^0|
   +|q_\alpha(t)-q_\alpha^0|+|q_\beta(t)-q_\beta^0|
   \le C_2\eps^{-1}+2C_vT\sqrt{\eps}\eps^{-3/2}
   =C^\ast\eps^{-1}, \]
with $C^\ast=C_2+2C_vT$. We remark that for the estimates
in this section the smallness of the $e_\alpha$ was not needed.

\subsection{Bounding $|\dot{v}_\alpha(t)|$}

Since
\[ \frac{d}{dt}\Big(m_{{\rm b}\alpha}\gamma_\alpha v_\alpha(t)\Big)
   =m_{0\alpha}(v_\alpha(t))\dot{v}_\alpha(t), \]
with the $(3\times 3)$-matrices $m_{0\alpha}(v_\alpha)$ given through
$m_{0\alpha}(v_\alpha)(z)=m_{{\rm b}\alpha}(\gamma_\alpha z
+\gamma_\alpha^3 (v_\alpha\cdot z)v_\alpha)$, $z\in\R^3$,
we obtain from (\ref{system2}) that for $\alpha=1, \ldots, N$
\begin{eqnarray}\label{dotv-1}
   \dot{v}_\alpha & = & {m_{0\alpha}(v_\alpha)}^{-1}
   \int d^3x\,\rho_\alpha(x-q_\alpha)\Big([E(x)-E_{v_\alpha}(x-q_\alpha)]
   +v_\alpha\wedge [B(x)-B_{v_\alpha}(x-q_\alpha)]\Big) \nonumber\\
   & = & {m_{0\alpha}(v_\alpha)}^{-1}
   \int d^3x\,\rho_\alpha(x)\Big(Z_1(x+q_\alpha, t)
   +v_\alpha\wedge Z_2(x+q_\alpha, t)\Big) + R_\alpha (t),
\end{eqnarray}
where ${m_{0\alpha}(v_\alpha)}^{-1}z={m_{{\rm b}\alpha}}^{-1}
\gamma_\alpha^{-1}(z-(v_\alpha\cdot z)v_\alpha)$, $z\in\R^3$,
is the matrix inverse of $m_{0\alpha}(v_\alpha)$. For (\ref{dotv-1})
it is important to note that adding the $E_{v_\alpha}(x-q_\alpha)$-term
and the $v_\alpha\wedge B_{v_\alpha}(x-q_\alpha)$-term does not change
the integral, as may be seen through Fourier transform using (\ref{EBv-def})
and (\ref{phiv-def}). Moreover, in (\ref{dotv-1}) we have set
\begin{equation}\label{R-def}
   R_\alpha(t)={m_{0\alpha}(v_\alpha)}^{-1}
   \bigg(\sum_{\stackrel{\beta=1}{\beta\neq\alpha}}^N
   \int d^3x\,\rho_\alpha(x-q_\alpha)
   \Big[E_{v_\beta}(x-q_\beta)+v_\alpha\wedge
   B_{v_\beta}(x-q_\beta)\Big]\bigg)
\end{equation}
and
\[ Z(x, t)=\left(\begin{array}{c} Z_1(x, t) \\ Z_2(x, t)
   \end{array}\right)
   =\displaystyle\left(\begin{array}{c} E(x, t)-\sum_{\beta=1}^N
   E_{v_\beta(t)}(x-q_\beta(t)) \\[1ex]
   B(x, t)-\sum_{\beta=1}^N B_{v_\beta(t)}(x-q_\beta(t))\end{array}\right). \]
Maxwell's equations and the relations
$(v \cdot\nabla) E_v(x) =-\nabla\wedge B_v(x)+e\varphi(x)v$,
$(v \cdot \nabla) B_v(x) =\nabla\wedge E_v(x)$, $e=e_\alpha$
for index $\alpha$, yield
\begin{equation}\label{Z-gleich}
   \dot{Z}(t)={\cal A}Z(t)-f(t)\,,\quad\mbox{with}\quad
   {\cal A}=\left(\begin{array}{cc} 0 & \nabla\wedge
   \\ -\nabla\wedge & 0 \end{array}\right)
\end{equation}
and
\begin{equation}\label{f-def}
   f(x, t)=\sum_{\beta=1}^N\left(\begin{array}{c}
   (\dot{v}_\beta(t)\cdot \nabla_v) E_{v_\beta}(x-q_\beta(t))
   \\[1ex] (\dot{v}_\beta(t)\cdot \nabla_v)
   B_{v_\beta}(x-q_\beta(t))\end{array}\right).
\end{equation}
The Maxwell operator ${\cal A}$ generates a $C^0$-group $U(t)$,
$t\in\R$, of isometries in $L^2(\R^3)^3\oplus L^2(\R^3)^3$; see
\cite[p.~435; (H2)]{dau-li}. Therefore we have the mild solution
representation
\begin{equation}\label{Z-form}
   Z(x, t)=[U(t)Z(\cdot, 0)](x)-\int_0^t ds\,[U(t-s)f(\cdot, s)](x).
\end{equation}
According to (\ref{ini-bed}), $Z(0)=0$, so the first term drops out.
To estimate the remaining term, we first state and prove some auxiliary
lemmas that will be used frequently.

\begin{lemma}\label{darst} For given $f=(f_1, f_2)$ with $\nabla\cdot f_1=0$
and $\nabla\cdot f_2=0$ we have for $W(t, s, x)=(W_1(t, s, x), W_2(t, s, x))
=[U(t-s)f(\cdot, s)](x)$
\begin{eqnarray*}
   W_1(t, s, x) & = & \frac{1}{4\pi (t-s)^2}\,\int_{|y-x|=(t-s)}\hspace{-1em}
   d^2y\,\Big[ (t-s)\nabla\wedge f_2(y, s)+f_1(y, s)+((y-x)\cdot\nabla)
   f_1(y, s)\Big], \\
   W_2(t, s, x) & = & \frac{1}{4\pi (t-s)^2}\,\int_{|y-x|=(t-s)}\hspace{-1em}
   d^2y\,\Big[ -(t-s)\nabla\wedge f_1(y, s)+f_2(y, s)+((y-x)\cdot\nabla)
   f_2(y, s)\Big].
\end{eqnarray*}
\end{lemma}
{\bf Proof\,:} See \cite[Lemma 8.1]{MK-S-2}.
{\hfill $\Box$}\bigskip

\begin{lemma}\label{max-esti} (a) Let $\xi(s)\ge 0$ be some function.
Assume that for $y\in\R^3$, $s\in [0, t]$, and some $f(y, s)
=(f_1(y, s), f_2(y, s))$ with $\nabla\cdot f_1=0=\nabla\cdot f_2$
\begin{eqnarray}
   |f_1(y, s)|+|f_2(y, s)| & \le & C\xi(s)\sum_{\beta=1}^N
   \frac{1}{1+{|y-q_\beta(s)|}^2}, \label{saram1} \\
   |\nabla f_1(y, s)|+|\nabla f_2(y, s)| & \le &
   C\xi(s)\sum_{\beta=1}^N\frac{1}{1+{|y-q_\beta(s)|}^3},
   \label{saram2}
\end{eqnarray}
Then for each $\alpha=1, \ldots, N$, $t\in [0, T\eps^{-3/2}]$,
and $|x|\le R_\varphi$
\[ \bigg|\int_0^t ds\,[U(t-s)f(\cdot, s)](x+q_\alpha(t))\bigg|
   \le C\bigg(\sup_{s\in [0, t]}\xi(s)\bigg). \]

\noindent (b) Under the hypotheses of (a), if instead of
(\ref{saram1}) and (\ref{saram2}) it holds for fixed $1\le\alpha\le N$
that
\begin{eqnarray}
   |f_1(y, s)|+|f_2(y, s)| & \le & C\xi(s)
   \sum_{\stackrel{\beta=1}{\beta\neq\alpha}}^N
   \frac{1}{1+{|y-q_\beta(s)|}^3}, \label{torga1} \\
   |\nabla f_1(y, s)|+|\nabla f_2(y, s)| & \le &
   C\xi(s)\sum_{\stackrel{\beta=1}{\beta\neq\alpha}}^N
   \frac{1}{1+{|y-q_\beta(s)|}^4},
   \label{torga2}
\end{eqnarray}
then for $t\in [0, T\eps^{-3/2}]$ and $|x|\le R_\varphi$ we have even that
\[ \bigg|\int_0^t ds\,[U(t-s)f(\cdot, s)](x+q_\alpha(t))\bigg|
   \le C\bigg(\sup_{s\in [0, t]}\xi(s)\bigg)\eps. \]

\noindent (c) Let $\xi(\tau, s)\ge 0$ be some function.
Assume that for $y\in\R^3$, $\tau\in [0, t]$, $s\in [0, \tau]$,
and some $g(y, \tau, s)=(g_1(y, \tau, s), g_2(y, \tau, s))$
with $\nabla\cdot g_1=0=\nabla\cdot g_2$ that
\begin{eqnarray}
   |g_1(y, \tau, s)|+|g_2(y, \tau, s)| & \le & C\xi(\tau, s)
   \sum_{\alpha=1}^N\frac{1}{1+{|y-q_\alpha(s)|}^3}, \label{saram1g} \\
   |\nabla g_1(y, \tau, s)|+|\nabla g_2(y, \tau, s)| & \le &
   C\xi(\tau, s)\sum_{\alpha=1}^N\frac{1}{1+{|y-q_\alpha(s)|}^4}.
   \label{saram2g}
\end{eqnarray}
Then for each $\alpha=1, \ldots, N$, $t\in [0, T\eps^{-3/2}]$,
and $|x|\le R_\varphi$
\[ \bigg|\int_0^t d\tau\int_0^{\tau}ds\,
   [U(t-s)g(\cdot, \tau, s)](x+q_\alpha(t))\bigg|
   \le C\bigg(\sup_{(\tau, s)\in\Delta_t}\xi(\tau, s)\bigg), \]
where $\Delta_t=\{(\tau, s): \tau\in [0, t], s\in [0, \tau]\}$.
\medskip

\noindent
In (a)--(c), all constants $C$ on the right-hand sides
are independent of $\alpha$, $t$, and $x$.
\end{lemma}
{\bf Proof\,:} (a) Define $W$ as in Lemma \ref{darst}. We derive
the estimates with $W_1$. Fix $1\le\alpha\le N$, $t\in [0, T\eps^{-3/2}]$,
$s\in [0, t]$, and $|x|\le R_\varphi$. According to
Lemma \ref{darst}, (\ref{saram1}), and (\ref{saram2})
\[ |W_1(t, s, x+q_\alpha(t))|
   \le C\frac{\xi(s)}{(t-s)^2}\,\sum_{\beta=1}^N
   I_{\alpha\beta}^{(2)}(t, s, x), \]
with
\begin{equation}\label{Indef}
   I_{\alpha\beta}^{(n)}(t, s, x)
   =\int_{|y-x-q_\alpha(t)|=(t-s)}\hspace{-1em} d^2y\,
   \bigg[\frac{(t-s)}{1+{|y-q_\beta(s)|}^{n+1}}
   + \frac{1}{1+{|y-q_\beta(s)|}^n}\bigg].
\end{equation}
In the sum in (\ref{Indef}), with general $n\ge 2$, we first consider
the term $I_{\alpha\alpha}^{(n)}(t, s, x)$, i.e., the one with $\beta=\alpha$.
In this case according to (\ref{v-bound}),
$|y-q_\beta(s)|\ge |y-x-q_\alpha(t)|-|x|-|q_\alpha(t)-q_\alpha(s)|
\ge (t-s)-R_\varphi-C\sqrt{\eps}(t-s)\ge (t-s)/2-R_\varphi$ for $\eps$ small.
Therefore $|y-q_\beta(s)|\ge (t-s)/4$ for $s\le t-4R_\varphi$.
We hence obtain for $\beta=\alpha$ and $s\le t-4R_\varphi$
\begin{equation}\label{anar}
   I_{\alpha\alpha}^{(n)}(t, s, x)
   \le C\,\frac{{(t-s)}^2}{1+{(t-s)}^n}.
\end{equation}
On the other hand, for $s\in [t-4R_\varphi, t]$
\begin{eqnarray*}
   I_{\alpha\alpha}^{(n)}(t, s, x) & \le & C{(t-s)}^2[(t-s)+1]
   \le C{(t-s)}^2[4R_\varphi+1]\le C{(t-s)}^2\frac{1}{1+{(4R_\varphi)}^n}
   \\ & \le & C\,\frac{{(t-s)}^2}{1+{(t-s)}^n}.
\end{eqnarray*}
Hence (\ref{anar}) shows that the latter estimate holds for any
$s\in [0, t]$. Since
\[ \int_0^t \frac{ds}{1+(t-s)^2}\le C,\quad
   \int_0^t d\tau \int_0^{\tau} \frac{ds}{1+(t-s)^3}\le C, \]
the term with $\beta=\alpha$ will satisfy the claimed estimates
not only in (a), but also in (c).

Next we turn to deriving a bound for $I_{\alpha\beta}^{(2)}(t, s, x)$
with $\beta\neq\alpha$. First note that for some portion of the interval
$[0, t]$ the preceding argument applies again. For this, define
$t_0=4(R_\varphi+C^\ast\eps^{-1})$. Then for $s\le t-t_0$
we find by (\ref{v-bound}) for $\eps$ small that on the $y$-sphere
\begin{eqnarray*}
   |y-q_\beta(s)| & \ge & |y-x-q_\alpha(t)|-|x|-|q_\alpha(t)-q_\beta(s)|
   \\ & \ge & (t-s)-R_\varphi-|q_\alpha(t)-q_\beta(t)|-|q_\beta(t)-q_\beta(s)|
   \\ & \ge & (t-s)-R_\varphi-C^\ast\eps^{-1}-C\sqrt{\eps}(t-s)
   \ge (t-s)/2-R_\varphi-C^{\ast}\eps^{-1}\ge (t-s)/4.
\end{eqnarray*}
Therefore as in (\ref{anar}) for general $n\ge 2$
\begin{equation}\label{horu}
   I_{\alpha\beta}^{(n)}(t, s, x)\le C\,\frac{{(t-s)}^2}{1+{(t-s)}^n},\quad
   s\in [0, t-t_0],
\end{equation}
and it remains to estimate $I_{\alpha\beta}^{(2)}(t, s, x)$ for
$\beta\neq\alpha$ and $s\in [t-t_0, t]$. To do so, we note
that an explicit computation shows for $z_1, z_2\in\R^3$
and $\gamma\ge 0$
\begin{eqnarray}
   \int_{|y-z_1|=\gamma}\hspace{-1em} d^2y\,
   \frac{1}{(1+{|y-z_2|}^2)} & = & \frac{\pi\gamma}{|z_1-z_2|}\,
   \log\Bigg(\frac{1+{(\gamma+|z_1-z_2|)}^2}{1+{(\gamma-|z_1-z_2|)}^2}\Bigg)
   \nonumber \\ & = & \frac{\pi\gamma}{|z_1-z_2|}\,\log\Bigg(1+
   \frac{4\gamma |z_1-z_2|}{1+{(\gamma-|z_1-z_2|)}^2}\Bigg)
   \nonumber \\ & \le & \frac{4\pi\gamma^2}
   {1+{(\gamma-|z_1-z_2|)}^2}, \label{2int}
\end{eqnarray}
as $\log(1+A)\le A$ for $A\ge 0$. Similarly, for $n\ge 2$
\begin{eqnarray}
   \lefteqn{\int_{|y-z_1|=\gamma}\hspace{-1em} d^2y\,
   \frac{1}{(1+{|y-z_2|}^{n+1})}} \nonumber\\ & = & 2\pi\gamma^2\int_{-1}^1
   \frac{dr}{1+{\Big({|z_1-z_2|}^2+2\gamma |z_1-z_2|r+\gamma^2\Big)}^{(n+1)/2}}
   \nonumber \\ & \le & C\gamma^2\int_{-1}^1
   \frac{dr}{{\Big(1+{|z_1-z_2|}^2+2\gamma |z_1-z_2|r+\gamma^2\Big)}^{(n+1)/2}}
   \nonumber \\ & = & C_n\frac{\gamma}{|z_1-z_2|}
   \Bigg(\frac{1}{{\big[1+{(|z_1-z_2|-\gamma)}^2\big]}^{(n-1)/2}}
   -\frac{1}{{\big[1+{(|z_1-z_2|+\gamma)}^2\big]}^{(n-1)/2}}\Bigg)\label{nint1}.
\end{eqnarray}
So in particular
\begin{equation}\label{nint2}
   \int_{|y-z_1|=\gamma}\hspace{-1em} d^2y\,
   \frac{1}{(1+{|y-z_2|}^{n+1})}\le
   C\frac{\gamma}{|z_1-z_2|},\quad n\ge 2.
\end{equation}
Below we will also need some more refined estimates, and for this purpose
we note that according to (\ref{nint1}) also
\begin{equation}\label{nint3}
   \int_{|y-z_1|=\gamma}\hspace{-1em} d^2y\,
   \frac{1}{(1+{|y-z_2|}^3)}\le
   C\frac{\gamma^2}{1+{(|z_1-z_2|+\gamma)}^2}
   \le C\frac{\gamma^2}{{|z_1-z_2|}^2}.
\end{equation}
Analogously we obtain
\begin{equation}\label{nint4}
   \int_{|y-z_1|=\gamma}\hspace{-1em} d^2y\,
   \frac{1}{(1+{|y-z_2|}^4)}\le
   C\,\frac{1}{1+{(|z_1-z_2|-\gamma)}^2}\,
   \min\bigg\{1,\,\frac{\gamma^2}{{|z_1-z_2|}^2}\bigg\}.
\end{equation}
As to bound $I_{\alpha\beta}^{(2)}(t, s, x)$ for
$\beta\neq\alpha$ and $s\in [t-t_0, t]$
we then use (\ref{nint2}) and (\ref{2int}) with $z_1=x+q_\alpha(t)$,
$z_2=q_\beta(s)$, and $\gamma=t-s$ to obtain for $s\in [t-t_0, t]$
\begin{equation}\label{I2ab}
   I_{\alpha\beta}^{(2)}(t, s, x)\le
   C\,\bigg(\frac{{(t-s)}^2}{|x+q_\alpha(t)-q_\beta(s)|}
   + \frac{{(t-s)}^2}{1+{[(t-s)-|x+q_\alpha(t)-q_\beta(s)|]}^2}\bigg).
\end{equation}
Therefore by (\ref{horu}) and (\ref{I2ab})
\begin{eqnarray}
   \lefteqn{\int_0^t ds\,\frac{\xi(s)}{(t-s)^2}\,I_{\alpha\beta}^{(2)}
   (t, s, x)} \nonumber \\ & \le &
   \bigg(\sup_{s\in [0, t]}\xi(s)\bigg)\,\bigg(\int_0^{t-t_0}
   \frac{ds}{(t-s)^2}\,I_{\alpha\beta}^{(2)}(t, s, x)
   + \int_{t-t_0}^t \frac{ds}{(t-s)^2}\,I_{\alpha\beta}^{(2)}(t, s, x)
   \bigg) \nonumber \\ & \le &
   C\bigg(\sup_{s\in [0, t]}\xi(s)\bigg)\,\bigg(\int_0^{t-t_0}
   \frac{ds}{1+(t-s)^2}
   + \int_{t-t_0}^t \frac{ds}{|x+q_\alpha(t)-q_\beta(s)|}
   \nonumber \\ & & \hspace{8em} + \int_{t-t_0}^t
   \frac{ds}{1+{[(t-s)-|x+q_\alpha(t)-q_\beta(s)|]}^2}\bigg).
   \label{bmbf}
\end{eqnarray}
The first of the three integrals is bounded by a constant.
Concerning the second, we have
\[ |x+q_\alpha(t)-q_\beta(s)|\ge
   |q_\alpha(t)-q_\beta(t)|-|x|-|q_\beta(t)-q_\beta(s)|\ge
   C_\ast\eps^{-1}-R_\varphi-C\sqrt{\eps}\,(t-s) \]
by (\ref{diff-bound}) and (\ref{v-bound}). In the domain of
integration $[t-t_0, t]$ it holds that $t-s\le t_0\le
C\eps^{-1}$, whence
\begin{equation}\label{ack}
   |x+q_\alpha(t)-q_\beta(s)|\ge C_\ast\eps^{-1}-R_\varphi
   -C\eps^{-1/2}\ge (C_\ast/2)\eps^{-1},\quad s\in [t-t_0, t],
   \quad \beta\neq\alpha,\quad |x|\le R_\varphi,
\end{equation}
for $\eps$ small. Therefore the second integral can be bound by
$C\eps\int_{t-t_0}^t ds\le C\eps t_0\le C$.
To estimate the last integral $=:J$ on the
right-hand side of (\ref{bmbf}), we substitute $\theta=t-s$
to obtain
\begin{equation}\label{J-def}
   J=\int_0^{t_0}\frac{d\theta}{1+{[\theta-r(\theta)]}^2}
\end{equation}
with $r(\theta)=|x+q_\alpha(t)-q_\beta(t-\theta)|$. Observe
that $|\dot{r}(\theta)|\le |\dot{q}_\beta(t-\theta)|\le C\sqrt{\eps}$
by (\ref{v-bound}). Thus $\theta\mapsto\chi(\theta)=\theta-r(\theta)$
is strictly increasing, and we can substitute $\theta=\theta(\chi)$
to get
\[ J=\int_{\chi(0)}^{\chi(t_0)}\frac{d\chi}{1-\dot{r}(\theta)}
   \bigg(\frac{1}{1+\chi^2}\bigg)\le C\int_{\R}\frac{d\chi}{1+\chi^2}
   \le C. \]
Summarizing these estimates we obtain the bound claimed in part
(a) of the lemma. \medskip

\noindent
(b) Defining $I^{(n)}_{\alpha\beta}$ as in (\ref{Indef}), we need to show
\begin{equation}\label{rus}
   \int_0^t \frac{ds}{(t-s)^2}\,I^{(3)}_{\alpha\beta}(t, s, x)\le C\eps,\quad
   \beta\neq\alpha.
\end{equation}
By (\ref{horu}),
\[ \int_0^{t-t_0}\frac{ds}{(t-s)^2}
   \,I_{\alpha\beta}^{(3)}(t, s, x)\le C\,
   \int_0^{t-t_0}\frac{ds}{(t-s)^2}
   \frac{(t-s)}{1+{(t-s)}^2}. \]
In the domain of integration, $(t-s)\ge t_0\le C\eps^{-1}$,
and hence
\begin{equation}\label{pors1}
   \int_0^{t-t_0}\frac{ds}{(t-s)^2}
   \,I_{\alpha\beta}^{(3)}(t, s, x)\le C\eps\int_0^t \frac{ds}{1+{(t-s)}^2}
   \le C\eps.
\end{equation}
Thus it remains to estimate the part of the integral in (\ref{rus})
for $s\in [t-t_0, t]$. Firstly, by (\ref{nint3}),
\begin{eqnarray}\label{pors2}
   \lefteqn{\int_{t-t_0}^t\frac{ds}{(t-s)^2}
   \int_{|y-x-q_\alpha(t)|=(t-s)}\hspace{-1em} d^2y\,
   \frac{1}{(1+{|y-q_\beta(s)|}^3)}} \nonumber\\ & & \le
   C\int_{t-t_0}^t\frac{ds}{|x+q_\alpha(t)-q_\beta(s)|^2}
   \le C\eps^2 \int_{t-t_0}^t\,ds=C\eps^2 t_0\le C\eps.
\end{eqnarray}
Here we have used $|x+q_\alpha(t)-q_\beta(s)|\ge
(C_\ast/2)\eps^{-1}$ for $\eps$ small, cf.~(\ref{ack}).
Reference to this is possible, since we again have that
$\beta\neq\alpha$. Analogously we infer from
(\ref{nint4}) that
\begin{eqnarray*}
   \lefteqn{\int_{t-t_0}^t\frac{ds}{(t-s)^2}
   \int_{|y-x-q_\alpha(t)|=(t-s)}\hspace{-1em} d^2y\,
   \frac{(t-s)}{(1+{|y-q_\beta(s)|}^4)}} \\ & \le &
   \int_{t-t_0}^t ds\,\frac{(t-s)}{{|x+q_\alpha(t)-q_\beta(s)|}^2}
   \,\bigg(\frac{1}{1+[(t-s)-|x+q_\alpha(t)-q_\beta(s)|]^2}\bigg)
   \\ & \le & C\eps^2 t_0\int_{t-t_0}^t\,
   \frac{ds}{1+[(t-s)-|x+q_\alpha(t)-q_\beta(s)|]^2}
   = C\eps J\le C\eps,
 \end{eqnarray*}
with the bounded $J$ from (\ref{J-def}). This together with
(\ref{pors2}) and (\ref{pors1}) shows that (\ref{rus}) is
satisfied. \medskip

\noindent
(c) Due to the remarks in (a), (\ref{saram1g}), and (\ref{saram2g})
we only have to prove
\begin{equation}\label{amson}
   \int_0^t d\tau\int_0^{\tau}\frac{ds}{(t-s)^2}\,
   I^{(3)}_{\alpha\beta}(t, s, x)\le C,\quad\beta\neq\alpha,\quad
   t\in [0, T\eps^{-3/2}],\quad |x|\le R_\varphi.
\end{equation}
We decompose the domain of integration $\Delta_t=\{(\tau, s): \tau\in [0, t],
s\in [0, \tau]\}$ in $\Delta_{t, 1}=\Delta_t\cap \{(\tau, s): s\in [0, t-t_0]\}$
and $\Delta_{t, 2}=\{(\tau, s): \tau\in [t-t_0, t], s\in [t-t_0, \tau]\}$.
On $\Delta_{t, 1}$ we can utilize (\ref{horu}) to
get
\begin{equation}\label{deli}
   \int\int_{\Delta_{t, 1}}d\tau ds\,\frac{1}{(t-s)^2}\,
   I^{(3)}_{\alpha\beta}(t, s, x)\le C \int_0^t d\tau \int_0^{\tau}ds
   \,\frac{1}{1+(t-s)^3}\le C.
\end{equation}
Since again $t-s\le t_0\le C\eps^{-1}$ for $(\tau, s)\in\Delta_{t, 2}$,
by (\ref{ack}) and (\ref{nint2})
\begin{eqnarray}\label{konr1}
   \lefteqn{\int\int_{\Delta_{t, 2}}d\tau ds\,\frac{1}{(t-s)^2}\,
   \int_{|y-x-q_\alpha(t)|=(t-s)}\frac{d^2y}{1+{|y-q_\beta(s)|}^3}}
   \nonumber\\ & \le & C\int\int_{\Delta_{t, 2}}d\tau ds\,\frac{1}{(t-s)}\,
   \frac{1}{|x+q_\alpha(t)-q_\beta(s)|}
   \nonumber\\ & \le & C\eps\int_{t-t_0}^t d\tau\int_{t-t_0}^{\tau}
   \frac{ds}{t-s}=C\eps\int_{t-}^t ds=C\eps t_0\le C.
\end{eqnarray}
In addition, by (\ref{nint4})
\begin{eqnarray}\label{konr2}
   \lefteqn{\int\int_{\Delta_{t, 2}}d\tau ds\,\frac{1}{(t-s)}\,
   \int_{|y-x-q_\alpha(t)|=(t-s)}\frac{d^2y}{1+{|y-q_\beta(s)|}^4}}
   \nonumber\\ & \le & \int\int_{\Delta_{t, 2}}d\tau ds\,\frac{1}{(t-s)}\,
   \frac{1}{1+{[(t-s)-|x+q_\alpha(t)-q_\beta(s)|]}^2}
   \nonumber \\ & = &
   \int_{t-t_0}^t \frac{ds}{1+{[(t-s)-|x+q_\alpha(t)-q_\beta(s)|]}^2}
   \le C,
\end{eqnarray}
since the last integral is just $J$ from (\ref{J-def}) and hence
bounded. By (\ref{deli}), (\ref{konr1}), and (\ref{konr2}) we thus have
proved (\ref{amson}).
{\hfill $\Box$}\bigskip

\begin{lemma}\label{phiv-esti} Define $\phi_v(x)$ through
$\hat{\phi}_v(k)=e\hat{\varphi}(k)/[k^2-(k\cdot v)^2]$.
Then for $x\in\R^3$ and $|v|\le\bar{v}<1$, with $\nabla=\nabla_x$,
\begin{eqnarray*}
   & |\nabla\phi_v(x)| + |\nabla_v\nabla\phi_v(x)|
   + |\nabla_v^2\nabla\phi_v(x)|
   + |\nabla_v^3\nabla\phi_v(x)|\le C|e|{(1+|x|)}^{-2}, & \\
   & |\nabla^2\phi_v(x)| + |\nabla_v\nabla^2\phi_v(x)| +
   |\nabla_v^2\nabla^2\phi_v(x)|
   + |\nabla_v^3\nabla^2\phi_v(x)|\le C|e|{(1+|x|)}^{-3}, & \\
   & |\nabla^3\phi_v(x)| + |\nabla_v\nabla^3\phi_v(x)| +
   |\nabla_v^2\nabla^3\phi_v(x)|
   + |\nabla_v^3\nabla^3\phi_v(x)|\le C|e|{(1+|x|)}^{-4}, & \\
   & |\nabla^4\phi_v(x)| + |\nabla_v\nabla^4\phi_v(x)| +
   |\nabla_v^2\nabla^4\phi_v(x)|
   + |\nabla_v^3\nabla^4\phi_v(x)|\le C|e|{(1+|x|)}^{-5}. &
\end{eqnarray*}
\end{lemma}
{\bf Proof\,:} Tedious calculations; see also the appendices of
\cite{MK-S-1, MK-S-2}.
{\hfill $\Box$}\bigskip

Now we can estimate $\int_0^t ds\,[U(t-s)f(\cdot, s)](x+q_\alpha(t))$,
cf.~(\ref{Z-form}), for $t\in [0, T\eps^{-3/2}]$ and $|x|\le R_\varphi$,
using Lemma \ref{darst} and Lemma \ref{max-esti}(a), with
$f=(f_1, f_2)$ defined by (\ref{f-def}). Since $\nabla\cdot B_v=0$,
and $\nabla\cdot E_v=e\varphi$ is independent of $v$,
we have $\nabla\cdot f_1=0=\nabla\cdot f_2$.
Concerning (\ref{saram1}) and (\ref{saram2}), note
$|\nabla_v E_v(x)|+|\nabla_v B_v(x)|\le
C(|\nabla\phi_v(x)|+|\nabla_v\nabla\phi_v(x)|)\le C|e|{(1+|x|)}^{-2}$ and
$|\nabla_v\nabla E_v(x)|+|\nabla_v\nabla B_v(x)|\le
C(|\nabla^2\phi_v(x)|+|\nabla_v\nabla^2\phi_v(x)|)\le C|e|{(1+|x|)}^{-3}$
by Lemma \ref{phiv-esti}. Thus (\ref{saram1}) and (\ref{saram2})
are satisfied with $\xi(s)=\Big(\max_{1\le\beta\le N}
|\dot{v}_\beta(s)|\Big)\Big(\max_{1\le\beta\le N}|e_\beta|\Big)$.
As $Z(x, 0)=0$, hence (\ref{Z-form}) in conjunction with
Lemma \ref{max-esti}(a) yields for $\alpha=1, \ldots, N$
\begin{equation}\label{gdhd}
   |Z(x+q_\alpha(t), t)|\le C\Big(\sup_{s\in [0, t]}\max_{1\le\beta\le N}
   |\dot{v}_\beta(s)|\Big)\Big(\max_{1\le\beta\le N}|e_\beta|\Big),\quad
   t\in [0, T\eps^{-3/2}],\quad |x|\le R_\varphi.
\end{equation}
We will utilize this further in (\ref{dotv-1}), and to this end we
also need to bound $R_\alpha(t)$ from (\ref{R-def}). For fixed
$\beta\neq\alpha$ one calculates for the interaction terms
\begin{eqnarray}
   \Psi_{\alpha\beta}(t) & = & \int d^3x\,\rho_\alpha(x-q_\alpha(t))
   \nabla\phi_{v_\beta(t)}(x-q_\beta(t)) \nonumber \\ & = &
   (-i)\,e_\alpha e_\beta\int d^3k\,k\frac{{|\hat{\varphi}(k)|}^2}
   {k^2-{(k\cdot v_\beta(t))}^2}\,e^{ik\cdot [q_\beta(t)-q_\alpha(t)]}
   \nonumber\\ & = &
   \frac{e_\alpha e_\beta}{4\pi}
   \int\int d^3x d^3y\,\varphi(x-q_\alpha(t))\varphi(y-q_\beta(t))
   \nabla\zeta_{v_\beta(t)}(x-y), \label{dxdy}
\end{eqnarray}
with
\begin{equation}\label{zeta-def}
   \zeta_v(x)=\frac{1}{{[(1-v^2)x^2+{(x\cdot v)}^2]}^{1/2}},\quad
   \hat{\zeta}_v(k)=\sqrt{\frac{2}{\pi}}\,\frac{1}{k^2-(k\cdot v)^2},
   \quad |v|<1.
\end{equation}
Then $\sup_{t\in [0, T\eps^{-3/2}]}|\nabla\zeta_{v_\beta(t)}(x)|
\le C{(1+|x|)}^{-2}$ due to (\ref{v-bound}). By $(C)$, in (\ref{dxdy})
we only need to integrate over $(x, y)$ that have
$|x-q_\alpha(t)|\le R_\varphi$ and $|y-q_\beta(t)|\le R_\varphi$.
Then by (\ref{diff-bound}), $|x-y|\ge |q_\alpha(t)-q_\beta(t)|
-2R_\varphi\ge C_\ast\eps^{-1}-2R_\varphi\ge (C_\ast/2)\eps^{-1}$
for $\eps$ small. Therefore (\ref{dxdy}) shows
\begin{equation}\label{Phi-esti}
   |\Psi_{\alpha\beta}(t)|\le C\eps^2,\quad t\in
   [0, T\eps^{-3/2}],\quad\alpha\neq\beta.
\end{equation}
By definition of $B_v(x)$ and $E_v(x)$ we have
\begin{equation}\label{ralph-form}
   R_\alpha(t)={m_{0\alpha}(v_\alpha(t))}^{-1}
   \sum_{\beta\neq\alpha}\Big(-\Psi_{\alpha\beta}(t)
   +[v_\beta(t)\cdot\Psi_{\alpha\beta}(t)]v_\beta(t)
   +v_\alpha(t)\wedge
   [-v_\beta(t)\wedge\Psi_{\alpha\beta}(t)]\Big)
\end{equation}
cf.~(\ref{R-def}), and therefore (\ref{Phi-esti}) together with
(\ref{v-bound}) implies
\begin{equation}\label{R-bound}
   |R_\alpha(t)|\le C\eps^2,\quad t\in [0, T\eps^{-3/2}].
\end{equation}
Hence (\ref{dotv-1}), (\ref{gdhd}), and (\ref{R-bound}) finally
yield
\[ |\dot{v}_\alpha(t)|\le C\Big(\sup_{s\in [0, t]}\max_{1\le\beta\le N}
   |\dot{v}_\beta(s)|\Big)\Big(\max_{1\le\beta\le N}|e_\beta|\Big)
   +C\eps^2, \]
for every $\alpha=1, \ldots, N$ and $t\in [0, T\eps^{-3/2}]$.
Choosing $\max_{1\le\beta\le N}|e_\beta|\le\bar{e}$ with
$\bar{e}>0$ sufficiently small, we find that for $\alpha=1, \ldots, N$
\begin{equation}\label{wstr}
    \sup_{t\in [0,\,T\eps^{-3/2}]}|\dot{v}_\alpha(t)|\le C\eps^2.
\end{equation}
For later reference we also note that then according to (\ref{gdhd})
\begin{equation}\label{Z-eps2}
   |Z(x+q_\alpha(t), t)|\le C\eps^2,\quad \alpha=1, \ldots, N,
   \quad t\in [0, T\eps^{-3/2}],\quad |x|\le R_\varphi.
\end{equation}

\subsection{Bounding $|\ddot{v}_\alpha(t)|$}

By (\ref{v-bound}) we have in particular that
\begin{equation}\label{v-diff-glob}
   |v_\alpha(t)-v_\beta(t)|\le C\sqrt{\eps},\quad t\in [0, T\eps^{-3/2}].
\end{equation}
In order to estimate the derivative of Equ.~(\ref{dotv-1}),
first note that using the explicit form of $m_{0\alpha}(v_\alpha)^{-1}$
we obtain from (\ref{wstr}) that
\begin{equation}\label{ddt-m0}
   \bigg|\frac{d}{dt}\,m_{0\alpha}(v_\alpha(t))^{-1}\bigg|
   \le C|\dot{v}_\alpha(t)|\le C\eps^2.
\end{equation}
Hence by (\ref{dotv-1}), (\ref{Z-eps2}) and (\ref{R-bound})
\begin{equation}\label{pach}
   |\ddot{v}_\alpha(t)|\le C\Big(\eps^4+|M_\alpha(t)|
   +|\dot{R}_\alpha(t)|\Big),
\end{equation}
with $R_\alpha$ defined in (\ref{R-def}), and
\begin{equation}\label{Malph-def}
   M_\alpha(t)=\int d^3x\,\rho_\alpha(x)\Big[(L_\alpha(t)Z_1)
   (x+q_\alpha(t), t)+v_\alpha(t)\wedge
   (L_\alpha(t)Z_2)(x+q_\alpha(t), t)\Big],
\end{equation}
where $L_\alpha(t)\phi=(v_\alpha(t)\cdot\nabla)\phi+\partial_t\phi$ for a
function $\phi=\phi(x, t)$. We first estimate $M_\alpha(t)$.
Let $\Sigma_\alpha(x, t)=(L_\alpha(t)Z)(x, t)$.
Since generally $\frac{d}{dt}[L_\alpha(t)\phi]=L_\alpha(t)\dot{\phi}
+(\dot{v}_\alpha\cdot\nabla)\phi$ and, see (\ref{Z-gleich}),
$\dot{Z}={\cal A}Z-f$ with $f$ from (\ref{f-def}), we obtain
\[ \dot{\Sigma}_\alpha={\cal A}\Sigma_\alpha
   +(\dot{v}_\alpha\cdot\nabla)Z-L_\alpha(t)f. \]
According to (\ref{ini-bed}) it may be shown that
$\Sigma_\alpha(x, 0)=0$. We hence get
\[ \Sigma_\alpha(x+q_\alpha(t), t)=\int_0^t d\tau\,\Big[U(t-\tau)
   \Big((\dot{v}_\alpha(\tau)\cdot\nabla)Z(\cdot, \tau)
   -L_\alpha(\tau)f(\cdot, \tau)\Big)\Big](x+q_\alpha(t)). \]
As a consequence of $\frac{d}{dt}(\nabla Z)=\nabla({\cal A}Z-f)
={\cal A}(\nabla Z)-\nabla f$ and $Z(x, 0)=0$, we obtain from the
group property of $U(\cdot)$ that
\begin{eqnarray*}
   \Sigma_{\alpha, 1}(x+q_\alpha(t), t) & = &
   \int_0^t d\tau\,\Big[U(t-\tau)\Big((\dot{v}_\alpha(\tau)\cdot\nabla)
   Z(\cdot, \tau)\Big)\Big](x+q_\alpha(t))
   \\ & = & -\int_0^t d\tau\int_0^{\tau}ds\,\Big[U(t-s)
   \Big((\dot{v}_\alpha(\tau)\cdot\nabla)
   f(\cdot, s)\Big)\Big](x+q_\alpha(t)).
\end{eqnarray*}
With $g(y, \tau, s)=\dot{v}_\alpha(\tau)\cdot\nabla f(y, s)$
it follows from the definitions of $f$, $E_v(x)$, and $B_v(x)$
that $\nabla\cdot g=0$. Moreover, by (\ref{wstr}) and Lemma \ref{phiv-esti}
we find that (\ref{saram1g}) and (\ref{saram2g}) are satisfied with
$\xi(\tau, s)=\eps^4$. Therefore Lemma \ref{max-esti}(c) applies
to yield for $\alpha=1, \ldots, N$
\begin{equation}\label{cpark}
   |\Sigma_{\alpha, 1}(x+q_\alpha(t), t)|\le C\eps^4,\quad
   t\in [0, T\eps^{-3/2}],\quad |x|\le R_\varphi.
\end{equation}
To estimate
\[ \Sigma_{\alpha, 2}(x+q_\alpha(t), t)
   =-\int_0^t d\tau\,\Big[U(t-\tau)
   \Big(L_\alpha(\tau)f(\cdot, \tau)\Big)\Big](x+q_\alpha(t)), \]
observe that
\begin{eqnarray*}
   [L_\alpha(\tau)f(\cdot, \tau)](x) & = &
   v_\alpha(\tau)\cdot\nabla f(x, \tau)+\partial_t f(x, \tau)
   \\ & = & \sum_{\beta=1}^N\Big\{
   (\ddot{v}_\beta\cdot\nabla_v)\Phi_{v_\beta}(x-q_\beta)
   +{(\dot{v}_\beta\cdot\nabla_v)}^2
   \Phi_{v_\beta}(x-q_\beta)\Big\}
   \\ & &  +\,\sum_{\stackrel{\beta=1}{\beta\neq\alpha}}^N
   \nabla_{xv}^2\Phi_{v_\beta}(x-q_\beta)(v_\alpha-v_\beta, \dot{v}_\beta)
   \\ & =: & f^\natural(\tau, y) + f^\flat(\tau, y),
\end{eqnarray*}
with all time arguments taken at time $\tau$, and $\Phi_v=(E_v, B_v)$.
Since $\nabla\cdot B_v=0$ and $\nabla\cdot E_v=e\varphi$ is independent of $v$,
we have that $\nabla\cdot f^\natural=0=\nabla\cdot f^\flat$. In addition,
$f^\natural$ satisfies (\ref{saram1}) and (\ref{saram2}) with
\[ \xi^\natural(\tau)=\Big(\max_{1\le\beta\le N}|
   \ddot{v}_\beta(\tau)|\Big)\Big(\max_{1\le\beta\le N}
   |e_\beta|\Big)+\eps^4. \]
Because $f^\flat$ has an additional $x$-derivative, moreover
(\ref{torga1}) and (\ref{torga2}) hold for $f^\flat$, with
\[ \xi^\flat(\tau)=\Big(\max_{1\le\beta\le N}
   |v_\alpha(\tau)-v_\beta(\tau)|\Big)\eps^2, \]
as again follows from Lemma \ref{phiv-esti} and (\ref{wstr}). Thus Lemma
\ref{max-esti}(a) and (b) imply that for all $\alpha=1, \ldots, N$,
$t\in [0, T\eps^{-3/2}]$, and $|x|\le R_\varphi$
\begin{eqnarray*}
   \lefteqn{|\Sigma_{\alpha, 2}(x+q_\alpha(t), t)|} \\ & \le &
   \bigg|\int_0^t d\tau\,[U(t-\tau)f^\natural(\cdot,
   \tau)](x+q_\alpha(t))\bigg| + \bigg|\int_0^t d\tau\,
   [U(t-\tau)f^\flat(\cdot, \tau)](x+q_\alpha(t))\bigg| \\ & \le &
   C\Big(\sup_{\tau\in [0, t]}\xi^\natural(\tau)
   +\sup_{\tau\in [0, t]}\xi^\flat(\tau)\eps\Big)
   \\ & \le & C\bigg[\eps^4+\Big(\sup_{\tau\in [0, t]}
   \max_{1\le\beta\le N}|\ddot{v}_\beta(\tau)|
   \Big)\Big(\max_{1\le\beta\le N}|e_\beta|\Big)
   +\Big(\sup_{\tau\in [0, t]}\max_{1\le\beta\le N}
   |v_\alpha(\tau)-v_\beta(\tau)|\Big)\eps^3\bigg].
\end{eqnarray*}
Hence by (\ref{cpark}) and (\ref{v-diff-glob}) for $\alpha=1, \ldots, N$,
$t\in [0, T\eps^{-3/2}]$, and $|x|\le R_\varphi$,
\[ |\Sigma_\alpha(x+q_\alpha(t), t)|\le
   C\bigg[\eps^{7/2}+\Big(\sup_{\tau\in [0, t]}
   \max_{1\le\beta\le N}|\ddot{v}_\beta(\tau)|\Big)
   \Big(\max_{1\le\beta\le N}|e_\beta|\Big)\bigg]. \]
According to the definition of $M_\alpha(t)$ in (\ref{Malph-def})
we therefore have
\begin{eqnarray}\label{stegl}
   |M_\alpha(t)| & = & \bigg|\int_{|x|\le R_\varphi} d^3x\,\rho_\alpha(x)
   \Big[\Sigma_{\alpha, 1}(x+q_\alpha(t), t)
   +v_\alpha(t)\wedge\Sigma_{\alpha, 2}(x+q_\alpha(t), t)\Big]\bigg|
   \nonumber\\ & \le &
   C\bigg[\eps^{7/2}+\Big(\sup_{\tau\in [0, t]}\max_{1\le\beta\le N}
   |\ddot{v}_\beta(\tau)|\Big)
   \Big(\max_{1\le\beta\le N}|e_\beta|\Big)\bigg].
\end{eqnarray}
To further estimate the right-hand side of (\ref{pach}), we have
to bound $\dot{R}_\alpha(t)$, with $R_\alpha(t)$ from
(\ref{R-def}). Calculating $\dot{R}_\alpha(t)$ explicitly we obtain
\begin{eqnarray*}
   \lefteqn{\dot{R}_\alpha(t)} \\ & = &
   \bigg(\frac{d}{dt}{m_{0\alpha}(v_\alpha)}^{-1}\bigg)
   m_{0\alpha}(v_\alpha)R_\alpha(t) \\ & &
   +\,{m_{0\alpha}(v_\alpha)}^{-1}
   \bigg(\sum_{\stackrel{\beta=1}{\beta\neq\alpha}}\int d^3x\,
   \rho_\alpha(x-q_\alpha)
   \Big[(\dot{v}_\beta\cdot\nabla_v) E_{v_\beta}(x-q_\beta)+v_\alpha\wedge
   (\dot{v}_\beta\cdot\nabla_v) B_{v_\beta}(x-q_\beta)\Big]\bigg) \\ & &
   +\,{m_{0\alpha}(v_\alpha)}^{-1}
   \bigg(\sum_{\stackrel{\beta=1}{\beta\neq\alpha}}
   \int d^3x\,\rho_\alpha(x-q_\alpha)
   \Big[((v_\alpha-v_\beta)\cdot\nabla)E_{v_\beta}(x-q_\beta) \\ & &
   \hspace{15.5em} + v_\alpha\wedge
   ((v_\alpha-v_\beta)\cdot\nabla)B_{v_\beta}(x-q_\beta)\Big]\bigg) \\ & &
   +\,{m_{0\alpha}(v_\alpha)}^{-1}
   \bigg(\sum_{\stackrel{\beta=1}{\beta\neq\alpha}}\int d^3x\,
   \rho_\alpha(x-q_\alpha)\,\dot{v}_\alpha\wedge B_{v_\beta}(x-q_\beta)\bigg)
   \\ & =: & \dot{R}_{\alpha, 1}(t) + \dot{R}_{\alpha, 2}(t)
   + \dot{R}_{\alpha, 3}(t) + \dot{R}_{\alpha, 4}(t)
\end{eqnarray*}
with all time arguments at time $t$. Firstly,
\begin{equation}\label{dotR1}
   |\dot{R}_{\alpha, 1}(t)|
   =\bigg|\bigg(\frac{d}{dt}{m_{0\alpha}(v_\alpha)}^{-1}\bigg)
   m_{0\alpha}(v_\alpha)R_\alpha(t)\bigg|\le C\eps^4
\end{equation}
for $\alpha=1, \ldots, N$ and $t\in [0, T\eps^{-3/2}]$ by
(\ref{ddt-m0}) and (\ref{R-bound}). Since $B_v(x)=-v\wedge\nabla\phi_v(x)$,
by (\ref{wstr}), (\ref{v-bound}), and (\ref{Phi-esti}) also
\begin{equation}\label{dotR2}
   |\dot{R}_{\alpha, 4}(t)|\le C\eps^{9/2}.
\end{equation}
What concerns $\dot{R}_{\alpha, 2}(t)$, we may repeat the
calculation in (\ref{dxdy}) to obtain
\begin{eqnarray*}
   \nabla_v\Psi_{\alpha\beta}(t) & := & \int d^3x\,\rho(x-q_\alpha(t))
   \nabla^2_{xv}\phi_{v_\beta(t)}(x-q_\beta(t)) \\ & = &
   \frac{1}{4\pi}\int\int d^3x d^3y\,\rho(x-q_\alpha(t))\rho(y-q_\beta(t))
   \nabla^2_{xv}\zeta_{v_\beta(t)}(x-y),
\end{eqnarray*}
with $\zeta_v(x)$ from (\ref{zeta-def}). Since
$\sup_{t\in [0,\,T\eps^{-3/2}]}|\nabla_{xv}^2\zeta_{v_\beta(t)}(x)|
\le C{(1+|x|)}^{-2}$, we get as before that
\[ |\nabla_v\Psi_{\alpha\beta}(t)|\le C\eps^2,
   \quad t\in [0, T\eps^{-3/2}],\quad \alpha\neq\beta, \]
and hence by (\ref{wstr})
\begin{equation}\label{dotR3}
   |\dot{R}_{\alpha, 2}(t)|\le C\eps^4.
\end{equation}
So finally we have to bound $\dot{R}_{\alpha, 3}(t)$,
and this relies on a similar argument. Here we have
\begin{eqnarray*}
   \nabla\Psi_{\alpha\beta}(t) & := & \int d^3x\,\rho(x-q_\alpha(t))
   \nabla^2\phi_{v_\beta(t)}(x-q_\beta(t)) \\ & = &
   \frac{1}{4\pi}\int\int d^3x d^3y\,\rho(x-q_\alpha(t))\rho(y-q_\beta(t))
   \nabla^2\zeta_{v_\beta(t)}(x-y),
\end{eqnarray*}
and $\sup_{t\in [0,\,T\eps^{-3/2}]}|\nabla^2\zeta_{v_\beta(t)}(x)|
\le C{(1+|x|)}^{-3}$. This in turn yields
\[ |\nabla\Psi_{\alpha\beta}(t)|\le C\eps^3,
   \quad t\in [0, T\eps^{-3/2}],\quad \alpha\neq\beta. \]
Using the explicit form of $E_v(x)$ and $B_v(x)$, as in
(\ref{ralph-form}), we then get for $t\in [0, T\eps^{-3/2}]$
\begin{equation}\label{dotR4}
   |\dot{R}_{\alpha, 3}(t)|\le C\eps^3\,|v_\alpha(t)-v_\beta(t)|
   \le C\eps^{7/2},
\end{equation}
by (\ref{v-diff-glob}). Summarizing (\ref{dotR1}), (\ref{dotR2}),
(\ref{dotR3}), and (\ref{dotR4}) it follows that
\begin{equation}\label{dotR-esti}
   |\dot{R}_\alpha(t)|\le C\eps^{7/2},\quad \alpha=1, \ldots, N,
   \quad t\in [0, T\eps^{-3/2}].
\end{equation}
Consequently, by (\ref{pach}), (\ref{stegl}), and (\ref{dotR-esti})
for $\alpha=1, \ldots, N$ and $t\in [0, T\eps^{-3/2}]$
\[ |\ddot{v}_\alpha(t)|\le C\Big(\eps^4+|M_\alpha(t)|+|\dot{R}_\alpha(t)|\Big)
   \le C\bigg[\eps^{7/2}+\Big(\sup_{\tau\in [0, t]}
   \max_{1\le\beta\le N}|\ddot{v}_\beta(\tau)|\Big)
   \Big(\max_{1\le\beta\le N}|e_\beta|\Big)\bigg]. \]
Choosing $\max_{1\le\beta\le N}|e_\beta|\le\bar{e}$ with
$\bar{e}$ sufficiently small we hence obtain
\[ \sup_{t\in [0,\,T\eps^{-3/2}]}|\ddot{v}_\alpha(t)|\le C\eps^{7/2},
   \quad\alpha=1, \ldots, N. \]
This completes the proof of Lemma \ref{esti}. {\hfill$\Box$}


\setcounter{equation}{0}

\section{Appendix B: Proof of Lemma \ref{tayl}}

Here we give the proof of Lemma \ref{tayl}. We verify e.g.~(b).
To compare the left-hand side to the right-hand side of the assertion,
we will insert some additional terms and estimate the corresponding
differences $D_j(t)$, $j=1, 2, 3$, for $t\in [t_0, T\eps^{-3/2}]$,
where $t_0=4(R_\varphi+C^\ast\eps^{-1})$. First we introduce
\begin{eqnarray}\label{D1-form}
   D_1(t) & = & i\int_0^t d\tau\int d^3k\,
   {|\hat{\varphi}(k)|}^2 e^{-ik\cdot\xi_{\alpha\beta}}
   \bigg\{ e^{-ik\cdot [q_\beta(t)-q_\beta(t-\tau)]}
   - e^{-ik\cdot [\tau v_\beta-\frac{1}{2}\tau^2\dot{v}_\beta]}\bigg\}\,
   \frac{\sin|k|\tau}{|k|}k \nonumber \\
   & = & -\,\nabla_\xi\int_0^t d\tau\int d^3k\,
   {|\hat{\varphi}(k)|}^2 e^{-ik\cdot\xi_{\alpha\beta}}
   \bigg\{ e^{-ik\cdot [q_\beta(t)-q_\beta(t-\tau)]}
   - e^{-ik\cdot [\tau v_\beta-\frac{1}{2}\tau^2\dot{v}_\beta]}\bigg\}\,
   \frac{\sin|k|\tau}{|k|} \nonumber \\
   & = & -\,\nabla_\xi\int\int d^3x d^3y\,\varphi(x)
   \varphi(y) \nonumber \\ & & \hspace{2em}\times\int_0^t
   d\tau\,\Big\{\psi_{\tau}\Big([\xi_{\alpha\beta}+x-q_\beta(t-\tau)]
   -[y-q_\beta(t)]\Big)-\psi_{\tau}\Big([x-\frac{1}{2}\tau^2 \dot{v}_\beta]
   -[y-\tau v_\beta]\Big)\Big\}, \nonumber \\
\end{eqnarray}
as follows through application of the Fourier transform,
with $\xi_{\alpha\beta}=q_\alpha(t)-q_\beta(t)$, and $\psi_\tau(x)=
{(4\pi|x|)}^{-1}$ for $|x|=\tau$ whereas $\psi_\tau(x)=0$ otherwise.
We claim that for $x, y\in\R^3$ with $|x|, |y|\le R_\varphi$
and $t\in [t_0, T\eps^{-3/2}]$ there exists a unique
$\tau_0=\tau_0(x, y, t, \xi_{\alpha\beta})\in [0, t_0]
\subset [0, t]$ such that
\begin{equation}\label{byn}
   \tau_0=\Big|[\xi_{\alpha\beta}+x-q_\beta(t-\tau_0)]
   -[y-q_\beta(t)]\Big|.
\end{equation}
To see this, observe with $\theta(\tau)
=\tau-|[\xi_{\alpha\beta}+x-q_\beta(t-\tau)]
-[y-q_\beta(t)]|$ that $0\ge\theta(0)\ge -(2R_\varphi+C^\ast\eps^{-1})$
and $\theta'(\tau)\ge 1-C_v\sqrt{\eps}$ by (\ref{diff-bound})
and (\ref{v-bound}). For $\eps$ so small that $1-C_v\sqrt{\eps}
\ge 1/2$ we hence obtain $\theta(t_0)\ge -(2R_\varphi+C^\ast\eps^{-1})
+t_0/2=2C^\ast\eps^{-1}$. This shows $\theta$ has a unique
zero $\tau_0\in [0, t_0]$. Moreover (\ref{byn}) together with
(\ref{diff-bound}) implies
\[ \tau_0\ge |\xi_{\alpha\beta}|-|x-q_\beta(t-\tau_0)]-[y-q_\beta(t)]|
   \ge C_\ast\eps^{-1}-2R_\varphi-C_v\sqrt{\eps}\tau_0, \]
whence also $\tau_0\ge C\eps^{-1}$ for $\eps$ small. Similarly,
we find a unique $\tau_1=\tau_1(x, y, t, \xi_{\alpha\beta})$ satisfying
\begin{equation}\label{tau1-def}   \
   \tau_1=\Big|[\xi_{\alpha\beta}+x-\frac{1}{2}\tau^2_1 \dot{v}_\beta]
   -[y-\tau_1 v_\beta]\Big|,
\end{equation}
with $\tau_1$ having the same properties as $\tau_0$. By definition
of $\psi_\tau$ we therefore may simply write
\begin{equation}\label{coln}
   D_1(t)=-\,\int\int d^3x d^3y\,\varphi(x)
   \varphi(y)\,\nabla_\xi\Big(\tau_0^{-1}-\tau_1^{-1}\Big).
\end{equation}
To estimate this, we calculate from (\ref{byn}) that
\begin{eqnarray*}
   \nabla_\xi\tau_0^{-1} & = & -\tau_0^{-3}\bigg\{
   \Big([\xi_{\alpha\beta}+x-q_\beta(t-\tau_0)]-[y-q_\beta(t)]\Big)
   \\ & & \hspace{3em}
   +\Big([\xi_{\alpha\beta}+x-q_\beta(t-\tau_0)]-[y-q_\beta(t)]\Big)\cdot
   v_\beta(t-\tau_0)\nabla_\xi\tau_0\bigg\},
\end{eqnarray*}
with an analogous expression for $\nabla_\xi\tau_1^{-1}$.
Therefore
\begin{eqnarray}\label{schle}
   \Big|\nabla_\xi\Big(\tau_0^{-1}-\tau_1^{-1}\Big)\Big|
   & \le & C\bigg(\tau_0^{-3}|q_\beta(t-\tau_0)-q_\beta(t-\tau_1)|
   \Big[1+|v_\beta(t-\tau_1)||\nabla_\xi\tau_1|\Big]
   \nonumber\\ & & \hspace{1.5em} +|\tau_0^{-3}-\tau_1^{-3}|\,
   \Big|[\xi_{\alpha\beta}+x-q_\beta(t-\tau_1)]-[y-q_\beta(t)]\Big|
   \Big[1+|v_\beta(t-\tau_1)||\nabla_\xi\tau_1|\Big]
   \nonumber \\ & & \hspace{1.5em}
   +\tau_0^{-2}|v_\beta(t-\tau_0)-v_\beta(t-\tau_1)|
   |\nabla_\xi\tau_1| \nonumber
   \\ & & \hspace{1.5em} +\tau_0^{-2}|v_\beta(t-\tau_0)|
   |\nabla_\xi(\tau_0-\tau_1)|\bigg).
\end{eqnarray}
>From (\ref{byn}), (\ref{tau1-def}), and according to the Taylor
expansion
\[ q_\beta(t-\tau)=q_\beta(t)-\tau v_\beta+\frac{1}{2}\tau^2\dot{v}_\beta
   + {\cal O}(\eps^{7/2}\tau^3), \]
cf.~Lemma \ref{esti}, it follows that
\begin{eqnarray*}
   |\tau_0-\tau_1| & \le & \Big|\tau_0
   v_\beta-\frac{1}{2}\tau_0^2\dot{v}_\beta-\tau_1 v_\beta
   +\frac{1}{2}\tau_1^2\dot{v}_\beta\Big|+{\cal O}(\eps^{7/2}\tau_0^3)
   \\ & \le & C\sqrt{\eps}\,|\tau_0-\tau_1|+C\eps^2(\tau_0+\tau_1)
   |\tau_0-\tau_1|+{\cal O}(\eps^{7/2}\tau_0^3),
\end{eqnarray*}
whence
\[ |\tau_0-\tau_1|={\cal O}(\sqrt{\eps}\,),\quad
   |\tau_0^{-3}-\tau_1^{-3}|={\cal O}(\eps^{9/2}), \]
recall $C\eps^{-1}\le\tau_0, \tau_1\le t_0={\cal O}(\eps^{-1})$.
Differentiating (\ref{byn}) and (\ref{tau1-def}) w.r.~to
$\xi=\xi_{\alpha\beta}$ we moreover get $|\nabla_\xi\tau_0|
+|\nabla_\xi\tau_1|={\cal O}(1)$, and after a longer calculation
which we omit also $|\nabla_\xi(\tau_0-\tau_1)|\le C\Big(\eps^{3/2}
+\sqrt{\eps}\,|\nabla_\xi(\tau_0-\tau_1)|\Big)$, thus
\[ |\nabla_\xi(\tau_0-\tau_1)|\le C\eps^{3/2}. \]
Utilizing these estimates and Lemma \ref{esti} in (\ref{schle}),
we consequently obtain $|\nabla_\xi(\tau_0^{-1}-\tau_1^{-1})|
\le C\eps^{7/2}$. Hence (\ref{coln}) yields
\begin{equation}\label{D1-esti}
   \sup_{t\in [t_0,\,T\eps^{-3/2}]}|D_1(t)|\le C\eps^{7/2}
\end{equation}
as desired. Next, with
\begin{eqnarray*}
   D_2(t) & = & i\int_0^t d\tau\int d^3k\,
   {|\hat{\varphi}(k)|}^2 e^{-ik\cdot \xi_{\alpha\beta}}
   \\ & & \hspace{2em} \times
   \bigg\{ e^{-ik\cdot [\tau v_\beta-\frac{1}{2}\tau^2\dot{v}_\beta]}
   -\Big(1-ik\cdot \Big[\tau v_\beta
   -\frac{1}{2}\tau^2\dot{v}_\beta\Big]-\frac{1}{2}\tau^2 {(k\cdot v_\beta)}^2
   \Big)\bigg\}\,\frac{\sin|k|\tau}{|k|}k
\end{eqnarray*}
it may be shown in a a similar way that
\begin{equation}\label{D2-esti}
   \sup_{t\in [t_0,\,T\eps^{-3/2}]}|D_2(t)|\le C\eps^{7/2}.
\end{equation}
Finally we need to compare $\int_0^t d\tau (\ldots)$
to the infinite $d\tau$-integral and thus let
\[ D_3(t)=i\int_t^{\infty} d\tau\int d^3k\,
   {|\hat{\varphi}(k)|}^2 e^{-ik\cdot \xi_{\alpha\beta}}
   \Big(1-ik\cdot \Big[\tau v_\beta
   -\frac{1}{2}\tau^2\dot{v}_\beta\Big]-\frac{1}{2}\tau^2
   {(k\cdot v_\beta)}^2\Big)\,\frac{\sin|k|\tau}{|k|}k. \]
With the notation
\[ K_p=e^{-ik\cdot\xi_{\alpha\beta}}\int_t^\infty d\tau\,
   \frac{\sin|k|\tau}{|k|}\,\tau^p,\quad p=0, \ldots, 2, \]
this may be rewritten as
\[ D_3(t)=\int d^3k\,{|\hat{\varphi}(k)|}^2\bigg(-\nabla_\xi K_0
   -(v_\beta\cdot\nabla_\xi)\nabla_\xi K_1+\frac{1}{2}(\dot{v}_\beta
   \cdot\nabla_\xi)\nabla_\xi K_2-\frac{1}{2}{(v_\beta\cdot\nabla_\xi)}^2
   \nabla_\xi K_2\bigg). \]
Thus we only need to estimate
\begin{eqnarray}\label{done}
   \int d^3k\,{|\hat{\varphi}(k)|}^2 K_p
   & = & \int d^3k\,{|\hat{\varphi}(k)|}^2
   e^{-ik\cdot\xi_{\alpha\beta}}\int_t^\infty d\tau\,
   \frac{\sin|k|\tau}{|k|}\,\tau^p \nonumber \\ & = &
   \int\int d^3x d^3y\,\varphi(x)
   \varphi(y)\,\int_t^\infty d\tau\,
   \psi_{\tau}(\xi_{\alpha\beta}+x-y)\,\tau^p,
\end{eqnarray}
the latter equality follows analogously to (\ref{D1-form}).
However, for $|x|, |y|\le R_\varphi$ and $t\in [t_0, T\eps^{-3/2}]$
we obtain in case $\tau=|\xi_{\alpha\beta}+x-y|$ from (\ref{diff-bound})
the contradiction
\[ 4(R_\varphi+C^\ast\eps^{-1})=t_0\le t\le\tau\le
   2R_\varphi+|\xi_{\alpha\beta}|\le 2R_\varphi
   +C^\ast\eps^{-1}. \]
This shows the term in (\ref{done}) is identically zero for
$t\in [t_0, T\eps^{-3/2}]$, and thus $D_3(t)=0$
for $t\in [t_0, T\eps^{-3/2}]$. Together with
(\ref{D1-esti}) and (\ref{D2-esti}) this completes
the proof of Lemma \ref{tayl}(b). {\hfill$\Box$}\bigskip

\noindent {\bf Acknowledgement:} We are grateful to
A.~Komech for discussions. HS thanks G. Sch\"{a}fer for useful hints 
on post-Newtonian corrections in general relativity and for insisting on 
(\ref{LD-dyn}).

\end{document}